\documentclass[showpacs,twocolumn,superscriptaddress,nofootinbib,10pt]{revtex4-2}
\usepackage{setspace}
\usepackage[utf8]{inputenc}
\usepackage{makecell}
\usepackage{float}
\usepackage{amsmath,amssymb,amsfonts,bm}
\usepackage{lipsum} 
\usepackage{graphicx}
\usepackage{multirow}
\usepackage{xcolor}
\definecolor{poscolor} {RGB} {252,188,190} 
\definecolor{negcolor} {RGB} {168,168,234} 
\usepackage{slashed}
\usepackage[normalem]{ulem}
\usepackage{braket}
\usepackage{booktabs} 
\usepackage{tabularx}

\usepackage{listings} 
\usepackage{amsmath}
\usepackage{txfonts}
\usepackage{amssymb}
\usepackage{slashed}
\usepackage{color}
\usepackage{multirow}
\usepackage{float}
\usepackage{graphicx,color,dcolumn}
\usepackage{subfigure}
\usepackage{orcidlink}
\usepackage{hyperref}
\hypersetup{colorlinks,citecolor=blue,anchorcolor=red,menucolor=red, linkcolor=red,filecolor=red,runcolor=red,urlcolor=blue,frenchlinks=true}
\usepackage{appendix}

\allowdisplaybreaks

\newcommand{\xju}{\affiliation{School of Physics Science and Technology, Xinjiang University, Urumqi, Xinjiang 830046 China}}

\begin{document}
\title{Role of the short-range dynamics in simultaneous interpretation of $P_{cs}$ pentaqurks via $\Xi_c^{(\prime,*)}\bar{D}^{(*)}$ molecules} 
\author{Ziye Wang\orcidlink{0009-0002-5536-6061}}
\email{zyewang@163.com}
\xju
\author{Nijiati Yalikun\orcidlink{0000-0002-3585-1863}}
\email{nijiati@xju.edu.cn}
\xju
\author{Yakefu Reyimuaji\orcidlink{0000-0002-4109-9110}}
\email{yreyi@hotmail.com}
\xju

\begin{abstract}
We investigate hidden-charm molecular states in $\Xi_c^{(\prime,*)}\bar{D}^{(*)}$ systems using the one-boson exchange model. By regulating the short-range interactions with parameter $a$ and cutoff $\Lambda$, we found ten bound states in isoscalar systems. Our analysis reveals that if the LHCb Collaboration's $P_{cs}(4459)$ and Belle Collaboration's $P_{cs}(4472)$ pentaqurks are indeed distinct states, their mass splitting can be resolved through $\Xi_c^{\prime}\bar{D}$-$\Xi_c\bar{D}^*$ coupled channel dynamics using consistent model parameters. This framework assigns $3/2^-$ and $1/2^-$ spin-parity quantum numbers to $P_{cs}(4459)$ and $P_{cs}(4472)$, respectively. With this consistent model parameter, we predict several new molecular candidates in the $4.3-4.7$ GeV mass region, demonstrating the crucial interplay between coupled channel effects and short-range dynamics in understanding hidden-charm pentaquarks as hadronic molecules. Additionally, we investigate the effects of $\Lambda\eta_c$ and $\Lambda J/\psi$ decay channels on the predicted molecular states, showing how these channels influence pole positions and provide insights into the detectability of these states through different production mechanisms.

\end{abstract}

\maketitle

\section{introduction}~\label{sec:1}
The formation mechanism of hadrons governed by the strong interaction in quantum chromodynamics (QCD) has been the subject of intense research. Nevertheless, understanding its low-energy dynamics remains one of the most formidable challenges in hadron physics. It is crucial to understand low energy QCD to determine whether exotic hadronic configurations exist beyond the well-established $qqq$ baryons and $q\bar{q}$ mesons of the conventional quark model~\cite{GellMann:1964nj,Zweig:1964jf}. These states, collectively referred to as exotic hadrons, containing multiquark states, glueballs, and quark-gluon hybrids. The multiquark sector comprises tetraquark ($qq\bar{q}\bar{q}$) and pentaquark ($qqqq\bar{q}$) configurations among other possible combinations; for recent reviews, see Refs.~\cite{Lebed:2016hpi,Esposito:2016noz,Guo:2017jvc,Liu:2019zoy,Chen:2022asf,Hanhart:2025bun,Wang:2025sic}. Elucidating the interaction of quarks within multiquark systems--particularly distinguishing between compact arrangements and loosely bound molecular structures--represents a fundamental challenge and main task in hadron community. 

In recent years, the LHCb Collaboration has observed many hidden-charm pentaquarks~\cite{LHCb:2015yax, LHCb:2019kea,LHCb:2020jpq, LHCb:2021chn, LHCb:2022ogu}. Most of them are located quite close to the thresholds of a pair of hadrons to which they can couple. This property can be understood as the presence of an $S$-wave attraction between the relevant hadron pair~\cite{Dong:2020hxe}, and it naturally leads to the hadronic molecule interpretation of the
pair~\cite{Chen:2016qju,Guo:2017jvc,Brambilla:2019esw,Yamaguchi:2019vea,Dong:2021juy,Dong:2021bvy}. Several models were proposed to explain the existence of these states as hadronic molecules even before the LHCb observations~\cite{Wu:2010jy,Wu:2010vk,Wu:2010rv,Wang:2011rga,Yang:2011wz,Wu:2012md,Xiao:2013yca,Karliner:2015ina}. The pentaquark states, $P_c(4450)$ and $P_c(4380)$, were observed by the LHCb Collaboration in 2015~\cite{LHCb:2015yax} . In their updated measurement~\cite{LHCb:2019kea}, the original $P_c(4450)$ state shows double peak structures identified as $P_c(4440)$ and $P_c(4457)$, but there is no clear evidence for the broad $P_c(4380)$. Meanwhile, a new narrow resonance $P_c(4312)$ showed up in this measurement. Numerous works set out to explain the three $P_c$ states simultaneously; see, e.g., Refs.~\cite{Liu:2019tjn,Xiao:2019aya,Du:2019pij,Du:2021fmf}. In the hidden charm with a strangeness sector, two $P_{cs}$ states were reported by the LHCb Collaboration, $P_{cs}(4459)$~\cite{LHCb:2020jpq} and $P_{cs}(4338)$~\cite{LHCb:2022ogu}, which are perfect candidates of $\Xi_c\bar D^*$ and $\Xi_c\bar D$ molecules, respectively~\cite{Dong:2021juy,Karliner:2022erb,Wang:2022mxy,Yan:2022wuz,Meng:2022wgl,Yang:2022ezl,Liu:2020hcv,Chen:2020kco,Wang:2020eep,Peng:2020hql,Chen:2020uif,Du:2021bgb,Xiao:2021rgp,Feijoo:2022rxf,Nakamura:2022jpd,Yan:2022wuz}. Very recently, the Belle and Belle II Collaborations added a likely member to the array
of hidden charm pentaquarks~\cite{Belle:2025pey}. They report a pentaquark state with the mass and width of $4471.7 \pm 4.8 \pm 0.6$ MeV and $21.9 \pm 13.1 \pm 2.7$ MeV, which is observed in $\Upsilon(1S,2S)\to J/\psi \Lambda \bar \Lambda$ decay at the $e^+e^-$ collider. We temporally denote it as $P_{cs}(4472)$ for the convenience of our future discussion. The mass of this state is about 13 MeV higher than that of the LHCb $P_{cs}(4459)$ pentaquark.  

The light meson exchange dynamics of the hadron interaction is the main mechanism used to explain the molecular structure of exotic hadrons, in which the interaction between hadrons are described by the one-boson exchange (OBE) model including the SU(3) vector-nonet mesons, pseudoscalar octet mesons, and the $\sigma$ meson as exchanged particles. The OBE model is quiet successful in  interpreting hadronic molecular pictures for hidden charm pentaquarks~\cite{He:2019rva,Chen:2019asm,Liu:2019zvb,Du:2021fmf,Yalikun:2021bfm,Yalikun:2023waw}. In the present work, we use the OBE model to investigate molecular states in $\Xi_c^{(\prime,*)}\bar D^{(*)}$ systems and explore the possibility of simultaneous interpretation in the $P_{cs}(4338)$ and $P_{cs}(4459)$ as well as $P_{cs}(4472)$ pentaquarks observed in recent experiments~\cite{LHCb:2020jpq, LHCb:2022ogu,Belle:2025pey}.  The OBE model has been used in pioneering works~\cite{Voloshin:1976ap,Tornqvist:1991ks} but suffers from a systematic expansion as offered by effective field theory approaches, where the short-range interaction is given in terms of contact terms with adjustable low-energy constants.

The characterization of short-range interactions provides critical insights into the formation mechanisms of the hadronic molecular states~\cite{Zou:2025sae}. Such short-range interactions in our work here can be represented with the $\delta(\bm r)$ term in coordinate space. In the OBE model, the potential may contain a $\delta(\bm r)$ term that acts as the short-range interactions. There are two strategies in the literature for handling this $\delta(\bm r)$ term: either retaining it~\cite{Liu:2009qhy,Wang:2020dya,Chen:2021tip,Wang:2022mxy,Wang:2024ukc} or discarding it~\cite{Thomas:2008ja,Liu:2019zvb,Ling:2021asz,Xu:2025mhc}. In our previous work \cite{Yalikun:2021bfm}, a parameter $a$ was introduced  to adjust the strength of the $\delta(\bm r)$ term. This parameter effectively introduces an additional contact interaction to account for extra short-range interactions from other heavier meson exchanges. As demonstrated in our previous study \cite{Yalikun:2021bfm}, a specific value of parameter $a$ enables the interpretation of the four observed $P_c$ states with a simultaneous cutoff. The analogous mechanism associated with the $\delta(\bm r)$ term within the $\Xi_c^{(\prime,*)}\bar D^{(*)}$ systems is the main content of the present work.

The paper is organized as follows.  The effective Lagrangian and OBE potentials for $\Xi_c^{(\prime,*)}\bar D^{(*)}$ systems as well as the treatment of the $\delta(\bm r)$ term in the OBE model are introduced in Sec.\ref{function}.  The terminology of the scattering matrix from the stationary Schr\"odinger equation and discussion on the dynamics of the OBE potentials in the possible $\Xi_c^{(\prime,*)}\bar D^{(*)}$ molecular states are given in Sec.\ref{Image Analysis}.  The summary of our results and conclusion  are presented in Sec.~\ref{sec:summary}

\section{effective potentials}~\label{sec:2}
\label{function}

In our study, we employ the OBE model to describe the interaction in $\Xi_c^{(\prime,*)}\bar D^{(*)}$ systems. To investigate the coupling between a charmed baryon or anti charmed meson with light scalar, pseudoscalar and vector mesons, we adopt the effective Lagrangian constructed with chiral symmetry and heavy quark spin symmetry \cite{Wise:1992hn, Falk:1992cx, Yan:1992gz, Casalbuoni:1996pg, Liu:2011xc, Yang:2011wz},
\begin{align}
	\mathcal{L}_{\rm{eff}}&=g_S\mathrm{Tr}[\bar{H}_a^{\bar{Q}}\sigma H_a^{\bar{Q}}]+ig\mathrm{Tr}[\bar{H}_a^{\bar{Q}}\gamma\cdot A_{ab}\gamma^5H_b^{\bar{Q}}]\notag\\
	&-i\beta\mathrm{Tr}[\bar{H}_a^{\bar{Q}}v_\mu(\Gamma_{ab}^\mu-\rho_{ab}^\mu)H_b^{\bar{Q}}]\notag\\
	&+i\lambda\mathrm{Tr}[\bar{H}_a^{\bar{Q}}\frac i2[\gamma_\mu,\gamma_\nu]F^{\mu\nu}(\rho_{ab})H_b^{\bar{Q}}\notag\\
	&+l_S\langle\bar{S}_\mu\sigma S^\mu\rangle-\frac32g_1\varepsilon^{\mu\nu\lambda\kappa}v_\kappa\langle\bar{S}_\mu A_\nu S_\lambda\rangle\notag\\
	&+i\beta_S\langle\bar{S}_\mu v_\alpha(\Gamma^\alpha-\rho^\alpha)S^\mu\rangle+\lambda_S\langle\bar{S}_\mu F^{\mu\nu}(\rho)S_\nu\rangle\notag\\
	&+i\beta_B\langle\bar{B}_{\bar{3}}v_\mu(\Gamma^\mu-\rho^\mu)B_{\bar{3}}\rangle+l_B\langle\bar{B}_{\bar{3}}\sigma B_{\bar{3}}\rangle\notag\\
	&+ ig_4\langle\bar{S}^\mu A_\mu B_{\bar{3}}\rangle+i\lambda_I\varepsilon^{\mu\nu\lambda\kappa}v_\mu\langle\bar{S}_\nu F_{\lambda\kappa}B_{\bar{3}}\rangle,\label{eq:lag}
\end{align}
where $\sigma$ represents the scalar meson field; Axial vector fields are $A_\mu=\frac12(\xi^\dagger\partial_\mu\xi-\xi\partial_\mu\xi^\dagger)=\frac i{f_\pi}\partial_\mu\mathbb{P}+\cdots,~\Gamma^\mu=\frac i2(\xi^\dagger\partial^\mu\xi+\xi\partial^\mu\xi^\dagger)=\frac i{2f_\pi^2}[\mathbb{P},\partial^\mu\mathbb{P}]+\cdots$, with $\xi=\exp(i\mathbb{P}/f_\pi)$ and $f_{\pi}=132$ MeV. The vector meson field is $\rho^\mu=\frac{ig_V}{\sqrt{2}}\mathbb{V}^{\mu}$, and the vector meson field strength tensor is $F^{\mu\nu}(\rho)=\partial^{\mu}\rho^{\nu}-\partial^{\nu}\rho^{\mu}+[\rho^{\mu},\rho^{\nu}]$. The pseudoscalar octet($\mathbb{P}$) and the vector nonet($\mathbb{V}$) mesons are written in the SU(3) matrix form as
\begin{align}\label{eq:meson-fields}
    \mathbb{P}&=\begin{pmatrix}
\frac{\eta}{\sqrt{6}} + \frac{\pi^0}{\sqrt{2}} & \pi^+ & K^+ \\
\pi^- & \frac{\eta}{\sqrt{6}} - \frac{\pi^0}{\sqrt{2}} & K^0 \\
K^- & \bar{K}^0 & -\sqrt{\frac{2}{3}} \eta
\end{pmatrix}, \\
\mathbb{V}&=\begin{pmatrix}
\frac{\rho^0}{\sqrt{2}} + \frac{\omega}{\sqrt{2}} & \rho^+ & K^{*+} \\
\rho^- & \frac{\omega}{\sqrt{2}} - \frac{\rho^0}{\sqrt{2}} & K^{*0} \\
K^{*-} & \bar{K}^{*0} & \phi
\end{pmatrix}. 
\end{align}
The $S$-wave heavy meson $\bar Q q$ and baryon $Qqq$ containing a single heavy quark can be represented with interpolated fields $H_a^{\bar Q}$ and $S^\mu$, respectively: 
\begin{align}
	H^{{\bar{Q}}}_a&=[\bar{P}_{a\mu}^{*}\gamma^{\mu}-\bar{P}_a\gamma_{5}]\frac{1-\slashed{v} }{2},\\
	\bar{H}_a^{\bar{Q}}&=\gamma^0H_a^{\bar{Q}\dagger}\gamma^0=\frac{1-\slashed{v}}2(\bar{P}_{a\mu}^{*\dagger}\gamma^\mu+\bar{P}_a^\dagger\gamma^5),\\
	S_\mu&=-\frac1{\sqrt{3}}(\gamma_\mu+v_\mu)\gamma^5B_{6}+B_{6\mu}^*,\\
	\bar{S}_\mu&=S_\mu^\dagger\gamma^0=\frac1{\sqrt{3}}\bar{B}_{6}\gamma^5(\gamma_\mu+v_\mu)+\bar{B}_{6\mu}^*.
\end{align}
where the scalar $\bar{P}_a$ and the vector $\bar{P}^*_{a\mu}$ fields are defined in the SU(3) flavor space as $(\bar{D}^0,\bar{D}^-,\bar{D}^0_s)$ and $(\bar{D}^{*0},\bar{D}^{*-},\bar{D}^{*0}_s)$ for $Q=c$. The charmed baryons with $J^P=1/2^+$ and $3/2^+$ in the $6_{\rm{F}}$ representation of SU(3) for the light quark flavor symmetry are labeled by $B_{6}$ and $B_{6}^{*,\mu}$, and written as
\begin{align}
B_{6}&=
\begin{pmatrix}
\Sigma_c^{++}&\Sigma_c^{+}/\sqrt 2&\Xi_c^{'+}/\sqrt 2\\
\Sigma_c^{+}/\sqrt 2&\Sigma_c^{0}&\Xi_c^{'0}/\sqrt 2\\
\Xi_c^{'+}/\sqrt 2&\Xi_c^{'0}/\sqrt 2&\Omega^{0}_c
\end{pmatrix},\\
B_{6}^{*}&=
\begin{pmatrix}
\Sigma_c^{*++}&\Sigma_c^{*+}/\sqrt 2&\Xi_c^{*+}/\sqrt 2\\
\Sigma_c^{*+}/\sqrt 2&\Sigma_c^{*0}&\Xi_c^{*0}/\sqrt 2\\
\Xi_c^{*+}/\sqrt 2&\Xi_c^{*0}/\sqrt 2&\Omega^{*0}_c
\end{pmatrix}.
\end{align}
The $S$-wave heavy baryons with $J^P=1/2^+$ in $\bar 3_{\rm{F}}$ representation are embedded in
\begin{align}
B_{\bar 3}&=\left(\begin{array}{ccc}
0&\Lambda_{c}^+&\Xi^+_{c}\\
-\Lambda_{c}^+&0&\Xi^0_{c}\\
-\Xi^+_{c}&-\Xi^0_{c}&0
\end{array}\right).
\end{align}

Through the expression of these elements, the above Lagrangian can be expanded to obtain the one Boson-exchange vertex. The specific form of the Lagrangian after expansion is described in detail in \cite{Chen:2020kco, Zhu:2021lhd, Wang:2023ael}. These effective Lagrangians will be used to derive the scattering amplitude, and thereby the OBE potential. By the Breit approximation \cite{Breit:1929zz, Breit:1930zza}, the scattering amplitude has the following relationship with the OBE potential in the momentum space:
\begin{equation}
	V^{{h_{1}h_{2}\to h_{3}h_{4}}}(\bm q)=-\frac{\mathcal{M}(h_{1}h_{2}\to h_{3}h_{4})}{\sqrt{2M_{1}2M_{2}2M_{3}2M_{4}}},
\end{equation}
where $M_i$ is the mass of the particle $h_i$, $\bm q$ is the three-momentum of the exchanged meson and $\mathcal{M}(h_1h_2\to h_3h_4)$ is the scattering amplitude of the transition $h_1h_2\to h_3 h_4$. In our derivation of the scattering amplitude, spinors of spin-$1/2$ and $3/2$ fermions with positive energy in nonrelativistic approximation read~\cite{Lu:2017dvm}
\begin{align}
u(p,m)_{B_{\bar 3 c}/B_{6 c}}&=\sqrt{2M_{B_{\bar 3 c}/B_{6 c}}}\begin{pmatrix}
\chi(m)\\0
\end{pmatrix},\\
u(p,m)_{B^*_{6 c}}^\mu&=\sqrt{2M_{B^*_{6 c}}}\begin{pmatrix}
(0,\bm\chi(m))^\mu\\(0,\bm 0)^\mu
\end{pmatrix},
\end{align} 
where $\chi(m)$ is the two-component spinor with third-component spin of $m$, and
\begin{align}
\bm{\chi}(m)=\sum_{m_1,m_2}\mathbb{C}_{1,m_1;1/2,m_2}^{3/2,m}\bm\epsilon(m_1)\chi(m_2),
\end{align} 
with $\bm\epsilon(\pm 1)=(\mp 1,-i,0)/\sqrt{2}$ and $\bm\epsilon(0)=(0,0,1)$. The scaled anti-heavy meson fields $\bar{P}$ and $\bar{P}^*$ are normalized as~\cite{Wise:1992hn,Wang:2020dya}
\begin{align}
   \langle 0 |\bar{P}| \bar c q(0^-)\rangle=\sqrt{M_{\bar{P}}},\quad    \langle 0|\bar{P}^*_\mu|\bar c q(1^-)\rangle=\epsilon_\mu\sqrt{M_{\bar{P}^*}}, 
\end{align} 
where $\epsilon_\mu$ is the polarization vector for the $\bar{P}^*_\mu$ field. 

The potential $V(\bm q)$ in the momentum space is transformed into the potential $V(\bm r)$ in the coordinate space by Fourier transform \cite{Chen:2016qju, Liu:2019zoy, Wang:2020dya}:
\begin{align}
	V(\bm r,\Lambda,m_{\rm{ex}})=\int\frac{\mathrm{d}^3 \bm q}{(2\pi)^3}V(\bm q)F^2(\bm q,\Lambda,m_{\rm{ex}})e^{iq\cdot r},
\end{align}
where $m_{\rm{ex}}$ is the mass of the exchange meson. $F^2(\bm q,\Lambda,m)$ is the form factor, which reduces the off-shell effects of the exchange meson and represents the internal structure of the interaction vertex \cite{Tornqvist:1993ng}. Based on the discussion of form factors in Ref.~\cite{Chen:2017vai}, the form factors generally take  monopole, dipole and exponential forms. In the case of low energy scale, the hadronic molecule should be almost unaffected by the type of form factor. In this study, we use the form factor in the monopole form  
\begin{align}
F(\bm q,\Lambda,m_{\rm{ex}})=\frac{\Lambda^2-m^2_{\rm{ex}}}{\Lambda^2+\bm q^2}.    
\end{align}

Since the structure of the interaction potential for $\Xi_c^{(\prime,*)}\bar D^{(*)}$ systems in the momentum space is quite similar to that in Ref.~\cite{Yalikun:2023waw}, we apply the results in the appendix A of this paper by replacing $D_s^{*}$ with $\bar D^{*}$. The momentum space potentials of $\Xi_c^{(\prime,*)}\bar D^{(*)}$ systems can be written in terms of three momentum space functions $1/(\bm q^2+m_{\rm{ex}}^2)$, $\bm A\cdot \bm q \bm B\cdot \bm q/(\bm q^2+m_{\rm{ex}}^2)$, and $(\bm A\times \bm q)\cdot(\bm B\times \bm q)/(\bm q^2+m_{\rm{ex}}^2)$, where $\bm A$ and $\bm B$ represent the spin operators. Therefore, their Fourier transformations are sufficient to write down all of the position space potentials for the $\Xi_c^{(\prime,*)}\bar D^{(*)}$ systems. The Fourier transformation of $1/(\bm q^2+m_{\rm{ex}}^2)$ is expressed as $Y_{\rm{ex}}$
\begin{equation}
	Y_{\rm{ex}}=\frac{1}{4\pi r}(e^{-m_{\rm{ex}}r}-e^{-\Lambda r})-\frac{\Lambda^{2}-m_{\rm{ex}}^{2}}{8\pi \Lambda}e^{-\Lambda r}.
	\label{eq:Yr}
\end{equation}

In Ref.~\cite{Yalikun:2023waw}, when analyzing the expression $\bm A \cdot \bm q \bm B \cdot \bm q / (\bm q^2 + m_{\rm{ex}}^2)$, it was demonstrated that the absence of a form factor leads to a Fourier transformation yielding a short-range $\delta(\bm r)$ potential in coordinate space.
With a form factor, this term dominates the short-range part of the potential. Refs.~\cite{Chen:2020kco, Wang:2023ael} discuss several hadronic molecules in $\Xi_c^{(\prime,*)}\bar D^{(*)}$ systems after removing the $\delta(\bm r)$ term. In this study, we analyze the role of the $\delta(\bm r)$ term in $\Xi_c^{(\prime,*)}\bar D^{(*)}$ systems. Here, we follow the results discussed in Refs.~\cite{Wang:2020dya, Yalikun:2021bfm, Yalikun:2023waw} regarding the $\delta(\bm r)$ term. The role of the $\delta(\bm r)$ term in the Fourier expansion of $\bm A\cdot \bm q \bm B\cdot \bm q/(\bm q^2+m_{\rm{ex}}^2)$ can be completely controlled once a dimensionless parameter $a$ is introduced,
\begin{align}
&\int\frac{d^3\bm q}{(2\pi)^3} \left (\frac{\bm A\cdot \bm q \bm B\cdot \bm q}{\bm q^2+m_{\rm{ex}}^2}-\frac{a}{3}\bm A\cdot \bm B\right ) \left (\frac{{\Lambda}^2-m^2_{\rm{ex}}}{\bm q^2+{\Lambda}^2}\right )^2 e^{i\bm q\cdot \bm r}\notag\\
&=-\frac{1}{3}[\bm A\cdot \bm BC_{\rm{ex}}+S(\bm A,\bm B,\hat r)T_{\rm{ex}}]\label{eq:CT-pot2},
\end{align}
where $S(\bm A,\bm B,\hat r)=3 \bm A\cdot \hat r \bm B\cdot \hat r-\bm A\cdot\bm B$ is the tensor operator in coordinate space, and the functions $C_{\rm{ex}}$ and $T_{\rm{ex}}$ read 
\begin{align}
C_{\rm{ex}}&=\frac{1}{r^2}\frac{\partial}{\partial r}r^2\frac{\partial}{\partial r}Y_{\rm{ex}}+\frac{a}{(2\pi)^3}\int \left (\frac{{\Lambda}^2-m^2_{\rm{ex}}}{\bm q^2+{\Lambda}^2}\right )^2 e^{i\bm q\cdot \bm r}d^3\bm q,\label{eq:Cr}\\
T_{\rm{ex}}&=r\frac{\partial}{\partial r}\frac{1}{r}\frac{\partial}{\partial r}Y_{\rm{ex}}.\label{eq:Tr}
 \end{align}   
 Apparently, the contribution of the $\delta(\bm r)$ term is fully included (excluded) when $a=0(1)$. Similarly, the Fourier transformation of the function $(\bm A\times \bm q) \cdot(\bm B\times \bm q)/(\bm q^2+m^2_{\rm {ex}})$ can be evaluated with the help of the relation $(\bm A\times \bm q) \cdot(\bm B\times \bm q)=\bm A\cdot \bm B|\bm q|^2-\bm A\cdot \bm q \bm B\cdot \bm q$. With these prescription, the coordinate space representations of the potentials can be written in terms of $Y_{\rm{ex}}$, $C_{\rm{ex}}$ and $T_{\rm{ex}}$ functions; they are collected in Appendix~\ref{sec:OBE-pot}.

 In this work, we focus on the negative parity states which are possibly bound in $S$ wave thus more easily form the molecular states compared to positive ones. The partial waves corresponding to the spin-parities of $J^P=1/2^{-},3/2^-,5/2^-$ are shown in Table~\ref{tab2}. To obtain corresponding potentials of these states, the position space potentials should be projected onto the partial waves listed in Table~\ref{tab2}, which is done by sandwiching the spin operators in the potentials between the partial waves of the initial and final states. We refer to Refs.~\cite{Yalikun:2021bfm,Yalikun:2021dpk} to compute the partial wave projections.
\begin{table*}[ht]
\caption{Partial waves of $J^P=1/2^{-},3/2^-$ and $5/2^-$ states in $\Xi_c^{(\prime,*)}\bar D^{(*)}$ systems}
\label{tab2}
\begin{ruledtabular}
\begin{tabular}{ccccccc}  
 & $\Xi_{c}\bar{D}$ & $\Xi_{c}'\bar{D}$ & $\Xi_{c}\bar{D}^{*}$ & $\Xi_{c}^{*}\bar{D}$ & $\Xi_{c}'\bar{D}^{*}$ & $\Xi_{c}^{*}\bar{D}^{*}$ \\ \hline
$J^{P}=\frac{1}{2}^{-}$ & $^{2}S_{\frac{1}{2}}$ & $^{2}S_{\frac{1}{2}}$ & $^{2}S_{\frac{1}{2}},~ ^{4}D_{\frac{1}{2}}$ & $^{4}D_{\frac{1}{2}}$ & $^{2}S_{\frac{1}{2}},~ ^{4}D_{\frac{1}{2}}$ & $^{2}S_{\frac{1}{2}},~ ^{4}D_{\frac{1}{2}} ,~ ^{6}D_{\frac{1}{2}}$ \\
$J^{P}=\frac{3}{2}^{-}$ & $^{2}D_{\frac{3}{2}}$ & $^{2}D_{\frac{3}{2}}$ & $^{4}S_{\frac{3}{2}},~ ^{2}D_{\frac{3}{2}},~ ^{4}D_{\frac{3}{2}}$ & $^{4}S_{\frac{3}{2}},~ ^{4}D_{\frac{3}{2}}$  & $^{4}S_{\frac{3}{2}},~ ^{2}D_{\frac{3}{2}},~ ^{4}D_{\frac{3}{2}}$ & $^{4}S_{\frac{3}{2}},~^{2}D_{\frac{3}{2}},~ ^{4}D_{\frac{3}{2}},~ ^{6}D_{\frac{3}{2}}$ \\ 
$J^{P}=\frac{5}{2}^{-}$ & $^{2}D_{\frac{5}{2}}$ & $^{2}D_{\frac{5}{2}}$ & $^{2}D_{\frac{5}{2}},~ ^{4}D_{\frac{5}{2}}$ & $^{4}D_{\frac{5}{2}}$ & $^{2}D_{\frac{5}{2}},~ ^{4}D_{\frac{5}{2}}$ & $^{6}S_{\frac{5}{2}},~^{2}D_{\frac{5}{2}},~ ^{4}D_{\frac{5}{2}},~ ^{6}D_{\frac{5}{2}}$ \\ 
\end{tabular}
\end{ruledtabular}
\end{table*}

\section{numerical analysis}~\label{sec:3}
\label{Image Analysis}

\subsection{The shape of the OBE potentials}\label{subsec:numerical-potentials}
In the previous section, we derived the OBE potentials of each channel in $\Xi_c^{(\prime,*)}\bar D^{(*)}$ systems. In this subsection, we discuss behavior of the various meson exchange potentials under the two extreme treatment of $\delta(\bm r)$ term. In the numerical analysis, we need to use the coupling constants of the Lagrangian and masses of the particles. For the coupling constants in the Lagrangian in Eq.~\eqref{eq:lag}, we adopt the values shown in Table \ref{coupling_constants}, where $\lambda$, $\lambda_S$, $\lambda_I$ are in units of GeV$^{-1}$. They are extracted from experimental data or deduced from various theoretical models~\cite{Ding:2008gr,Liu:2011xc,Meng:2019ilv,Isola:2003fh}. These coupling coefficients are in agreement with the ones in Ref.~\cite{Zhu:2021lhd} except for $g$ (the $\bar{D}^{*}\bar{D}^{*}\pi$ coupling),  and the negative value of $g$ is used to be consistent with the quark model. 
\begin{table}[ht]
\centering
\caption{Coupling constants.}
\label{coupling_constants}
\begin{ruledtabular}
\begin{tabular}{c c c c c c c c}
\(g_S\) & \(l_B\) & \(l_S\) & \(g\) & \(g_1\) & \(g_4\) &  \(\beta\) \\
0.76 & $-l_S/2$ & 6.2 & -0.59 & 0.94 & $3g_1/(2\sqrt{2})$ & 0.9 &  \\
\midrule
\(\beta_B\) &\(\beta_S\) & \(\lambda\) & \(\lambda_S\) & \(\lambda_I\) & \(g_v\) & &  \\
$-\beta_S/2$ & $-1.74$ & \(0.56\) & \(-3.31\) & $-\lambda_S/\sqrt{8}$ & 5.9 & & \\
\end{tabular}
\end{ruledtabular}
\end{table} 
For the masses of the exchanged mesons, we adopt the isospin averaged masses as $m_{\pi}=138.0$ MeV, $m_{\eta}=547.9$ MeV, $m_{\rho}=770.7$ MeV, $m_{\omega}=782.0$ MeV and $m_{\sigma}=600.0$ MeV, which are taken from the  Review of Particle Physics (RPP)~\cite{ParticleDataGroup:2024cfk}. 
The $\sigma$ meson refers to the lightest scalar meson with isospin $0$ and spin-parity $0^+$, corresponding to $f_0(500)$ in the RPP, which is a very broad state with large mass uncertainty ($400-800$ MeV). It has been shown that contributions from such a broad resonance exchange (effectively correlated scalar-isoscalar $2\pi$ exchange) in the $t$ channel can be represented by a stable particle with a mass of about 600 MeV in the nucleon-nucleon interaction~\cite{Meissner:1990kz,Wu:2023uva}. We simply set the mass of $\sigma$ to $600$ MeV, which is also commonly used in the OBE model for hadronic molecules~\cite{Liu:2009qhy,Liu:2019zvb,Xu:2025mhc,Yalikun:2025ssz}. 

\begin{figure}[htbp]
\centering 
\includegraphics[width=0.48\textwidth]{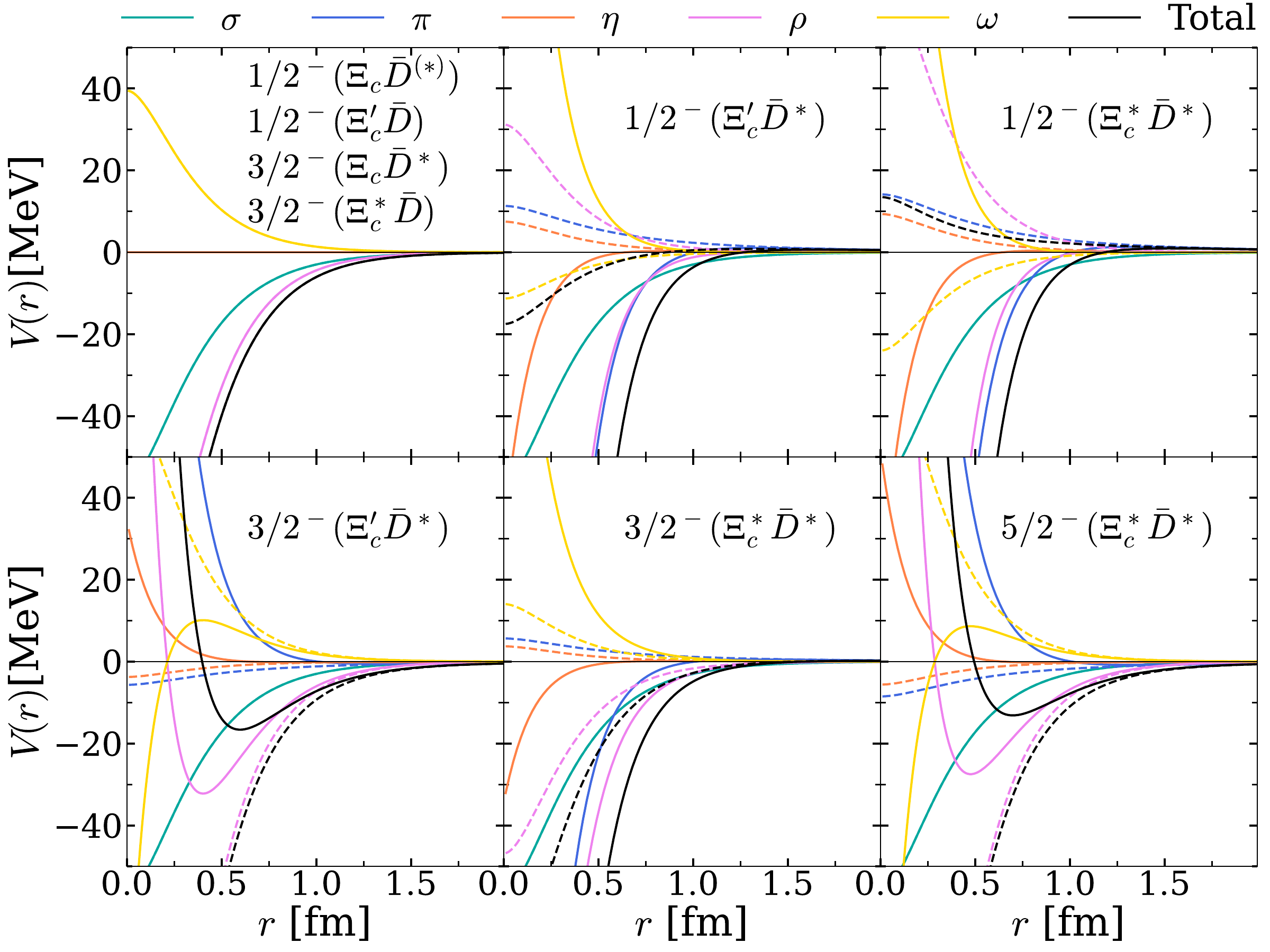} 
\caption{The $S$-wave OBE potentials of isoscalar systems for the $J^P=1/2^-, 3/2^-, 5/2^-$ states with $\Lambda= 1.2$ GeV. The solid and dashed curves are corresponding to the cases with $a=0$ and $1$, respectively.}
\label{V(r)_I=0}
\end{figure}

\begin{figure}[htbp]
	\centering 
	\includegraphics[width=0.48\textwidth]{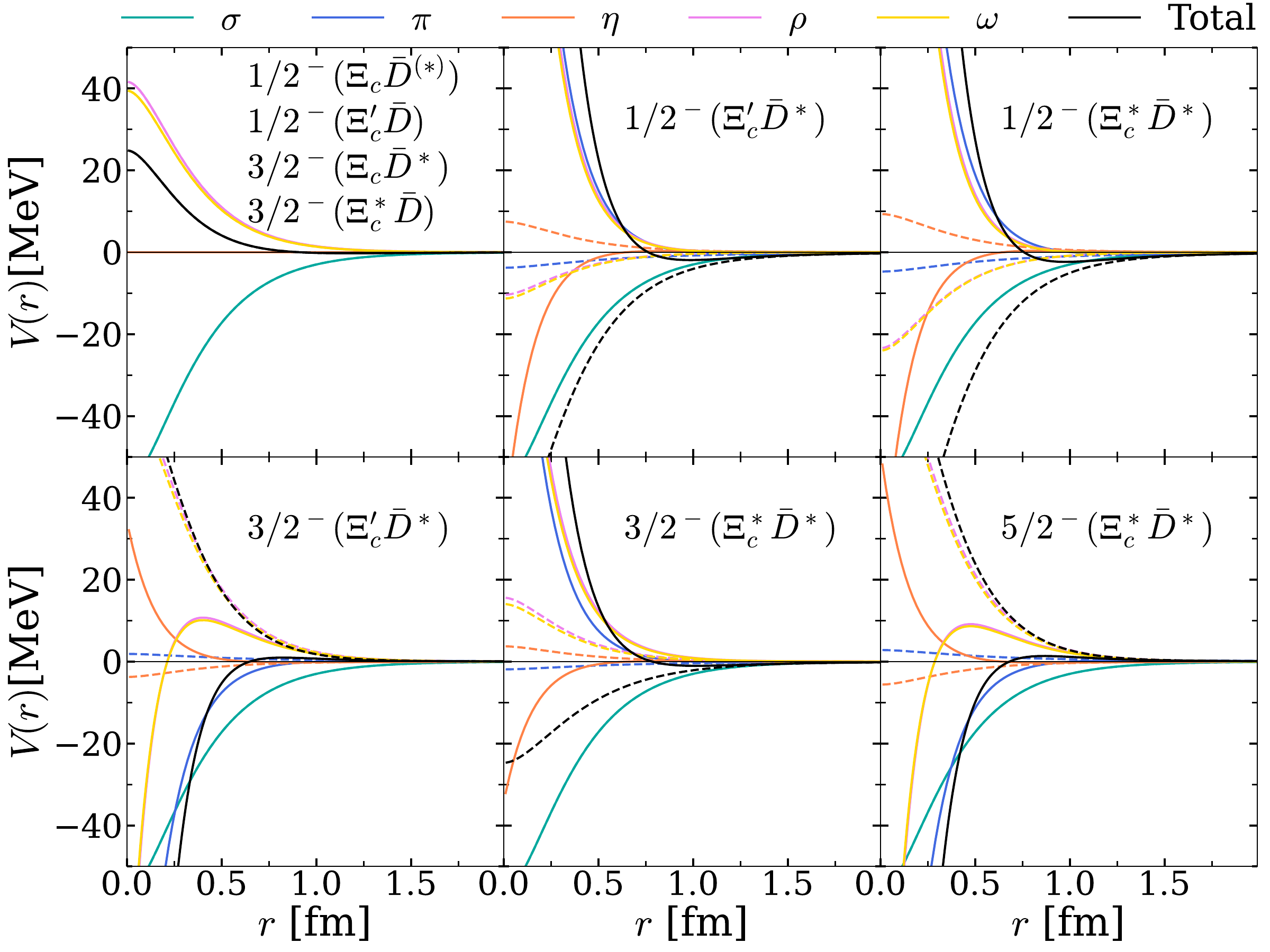} 
	\caption{Plots similar to Fig.~\ref{V(r)_I=0} for isovector system.} 
	\label{V(r)_I=1}
	\end{figure}

Since the contributions from other channels with higher partial waves are strongly suppressed by repulsive centrifugal terms, the potentials for S-wave channels are crucial for formation of hadronic molecules. Therefore, we present the OBE potentials for $S$-wave channels in $\Xi_c^{(\prime,*)}\bar D^{(*)}$ systems with $J^P=1/2^-, 3/2^-, 5/2^-$. The OBE potentials as functions of coordinate $r$ for isoscalar($I=0$) and isovector($I=1$) systems are shown in Figs.~\ref{V(r)_I=0} and \ref{V(r)_I=1}, respectively. In each subplot, the solid curves represent potentials with the $\delta(\bm{r})$ term included ($a=0$) while the dashed curves correspond to $a=1$ cases. Notably, the OBE potentials for the $1/2^-$ systems ($\Xi_c\bar D^{(*)}$ and $\Xi_c'\bar D$) and $3/2^-$ systems ($\Xi_c\bar D^*$ and $\Xi_c^*\bar D$) are identical in the non-relativistic limit. These potentials arise solely from scalar and vector meson exchanges, as dictated by the symmetry properties of the Lagrangian in Eq.~\eqref{eq:lag}. Importantly, the potentials of these five channels are independent of the $\delta(\bm{r})$ term since the contribution of the tensor coupling in corresponding $t$-channel amplitude is zero. In contrast, potentials for other channels depend on the $\delta(\bm{r})$ term.

In the OBE framework, the $\sigma$ meson exchange provides a spin-independent attractive potential that is unaffected by the $\delta(\bm{r})$ term, while contributions from other mesons depend on spin and the $a$ parameter. The isovector mesons ($\rho$, $\pi$) and isoscalar mesons ($\eta$, $\omega$) exhibit opposite potential signs in a specific isospin system. For $I=0$ systems, the $\eta$ and $\omega$ potentials counterbalance those of $\pi$ and $\rho$, whereas this relationship reverses for $I=1$ systems. The $1/2^-(\Xi_c'\bar{D}^*,\Xi_c^*\bar{D}^*)$ and $3/2^-(\Xi_c^*\bar{D}^*)$ channels display attractive potentials when $a=0$, and turn to weak or repulsive interactions at $a=1$. Conversely, the $3/2^-(\Xi_c'\bar{D}^*)$ and $5/2^-(\Xi_c^*\bar{D}^*)$ channels show repulsive potentials at $a=0$ that become attractive at $a=1$. This behavior inverts for $I=1$ systems as shown in Fig.~\ref{V(r)_I=1}. 

A critical observation is the sign reversal in vector/pseudoscalar meson exchange potentials near $r\sim0.5$ fm. This originates from the short-range $\delta(\bm{r})$ term's contribution, which opposes the sign of the remaining potential components. Such behavior underscores the non-trivial role of form factors in mediating meson exchange dynamics across different spatial ranges.

\subsection{Possible molecular states}
\label{bound energy}

The LHCb Collaboration first observed hidden-charm pentaquarks through analysis of $J/\psi p$ invariant mass distributions, with masses lying several MeV below the $\Sigma_c\bar{D}^{(*)}$ thresholds~\cite{LHCb:2015yax,LHCb:2019kea}. These states are formulated to emerge as molecular bound states of $\Sigma_c^{(*)}\bar{D}^{(*)}$ systems, where the interaction dynamics can be effectively described through non-relativistic potentials deduced from $t$-channel scattering amplitudes within the OBE framework~\cite{Chen:2019asm,Liu:2019tjn,Xiao:2019aya,Du:2019pij,Du:2021fmf,Yalikun:2021bfm}. 

In our analysis of $\Xi_c^{(\prime,*)}\bar{D}^{(*)}$ systems, we extend this formalism to include $S$-$D$ wave mixing effects through coupled channel calculations. The radial Schr\"odinger equation for the potential matrix $V_{jk}$ takes the form
\begin{equation}
\left[-\frac{1}{2\mu_j}\frac{d^2}{dr^2} + \frac{l_j(l_j+1)}{2\mu_jr^2} + W_j\right]u_j^{l_j} + \sum_k V_{jk}u_k^{l_j} = Eu_j^{l_j},\label{eq_schro_coupl}
\end{equation}
where $j$ denotes the channel index, $u_j^{l_j}(r) \equiv rR_j^{l_j}(r)$ represents the reduced radial wave function with angular momentum quantum number $l_j$, $\mu_j$ the reduced mass, and $W_j$ the threshold for channel $j$. The momentum for channel $j$ can be expressed as 
\begin{align}\label{eq:ch-mom}
q_j(E)=\sqrt{2\mu_j(E-W_j)}.
\end{align}
 By solving Eq.~\eqref{eq_schro_coupl}, we obtain the wave function that is normalized to satisfy the incoming boundary condition for the $j$th channel~\cite{osti_4661960},
\begin{eqnarray}
u_{jk}^{l_j}(r)\overset{r\rightarrow \infty}{\longrightarrow} \delta_{jk}h^-_{l_j}(q_j r)-S_{jk}(E)h^+_{l_j}(q_j r)\label{eq:asym-wave},
\end{eqnarray}
where $h_l^\pm(x)$ are spherical Hankel functions and $S_{jk}(E)$ is the scattering matrix component. Bound/virtual states and resonances are represented as the poles of $S_{jk}(E)$ in the complex energy plane~\cite{osti_4661960}. Note that channel momentum $q_j$ is a multivalued function of energy $E$, there are two Riemann sheets in the complex energy plane for each channel: one is called the first or physical sheet, while the other one is called the second or unphysical sheet. In the physical sheet, complex energy $E$ maps to the upper half plane ($\rm{Im}[q_j]\geq 0$) of the channel momentum $q_j$. In the unphysical sheet, complex energy $E$ maps to the lower half plane ($\rm{Im}[q_j]< 0$) of the channel momentum $q_j$. Poles may appear in this sheet, and those poles correspond to resonances if their real parts are larger than the thresholds of some channels. Bound state mass and resonance have the relation with pole position $E_{\mathrm{pole}}$  
\begin{align}
	E_{\mathrm{pole}}&=M,~~~~~~~~~~\text{( at physical sheet)}\\
    E_{\mathrm{pole}}&=M-i\Gamma/2,~~~\text{(at unphysical sheet)}
\end{align}
where $M$ is the mass of the bound/resonance state, and $\Gamma$ is the decay width.
This coupled channel formulation enables rigorous treatment of near-threshold states while preserving unitarity constraints~\cite{Yalikun:2021bfm}. 

\begin{figure}[htbp]
\centering 
\includegraphics[width=0.5\textwidth]{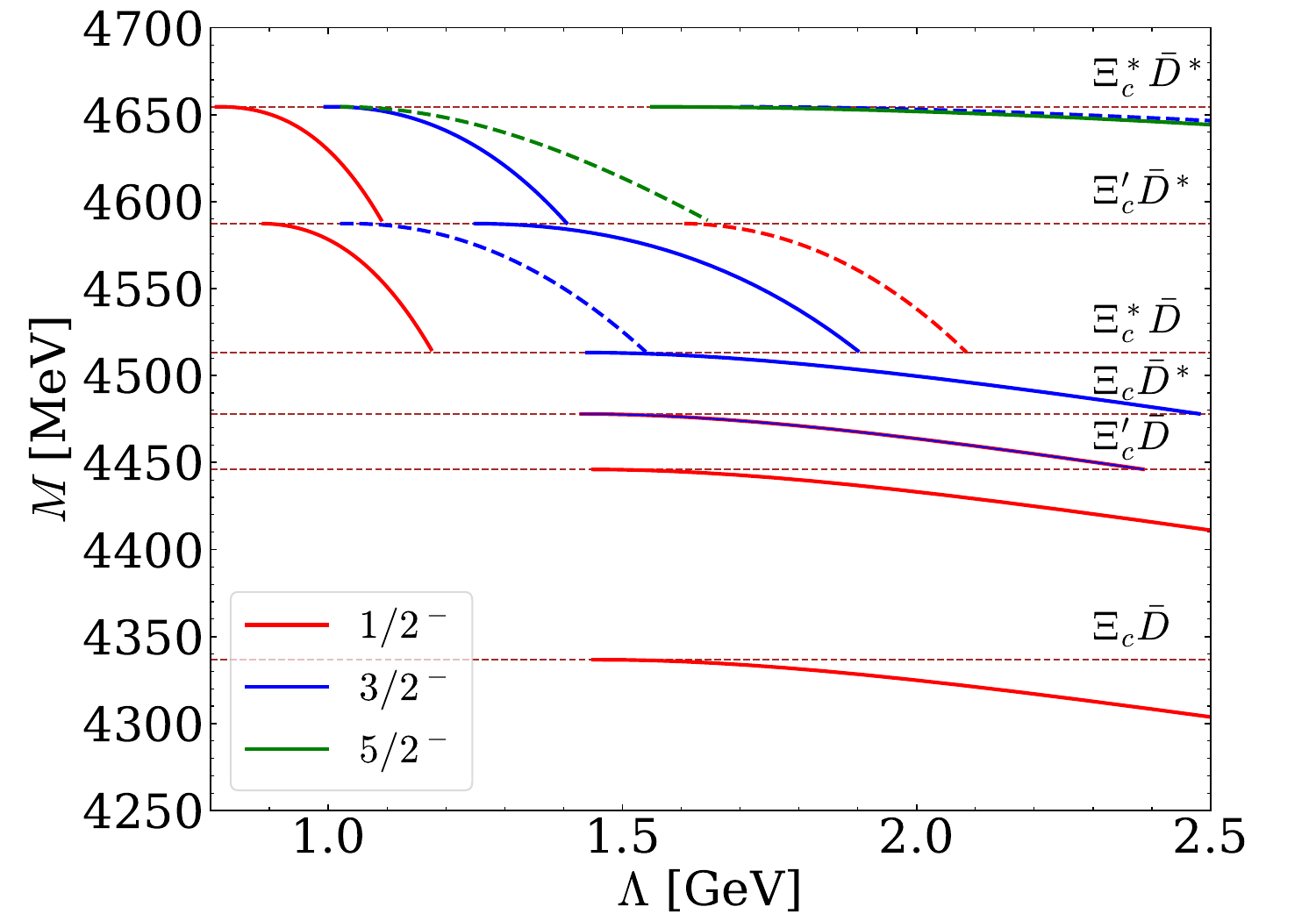}
\caption{Masses of the bound states with $I=0$ and $J^P=1/2^-, 3/2^-, 5/2^-$ quantum numbers in $\Xi_c^{(\prime,*)}\bar D^{(*)}$ systems . For curves of the same color, the solid(dashed) lines represent the $a=0(1)$ case. Curves for the $1/2^-$ and $3/2^-$ bound states in $\Xi_c\bar D^*$ channel are overlapped.}
\label{E(I=0)}
\end{figure}

\begin{figure}[htbp]
\centering 
\includegraphics[width=0.5\textwidth]{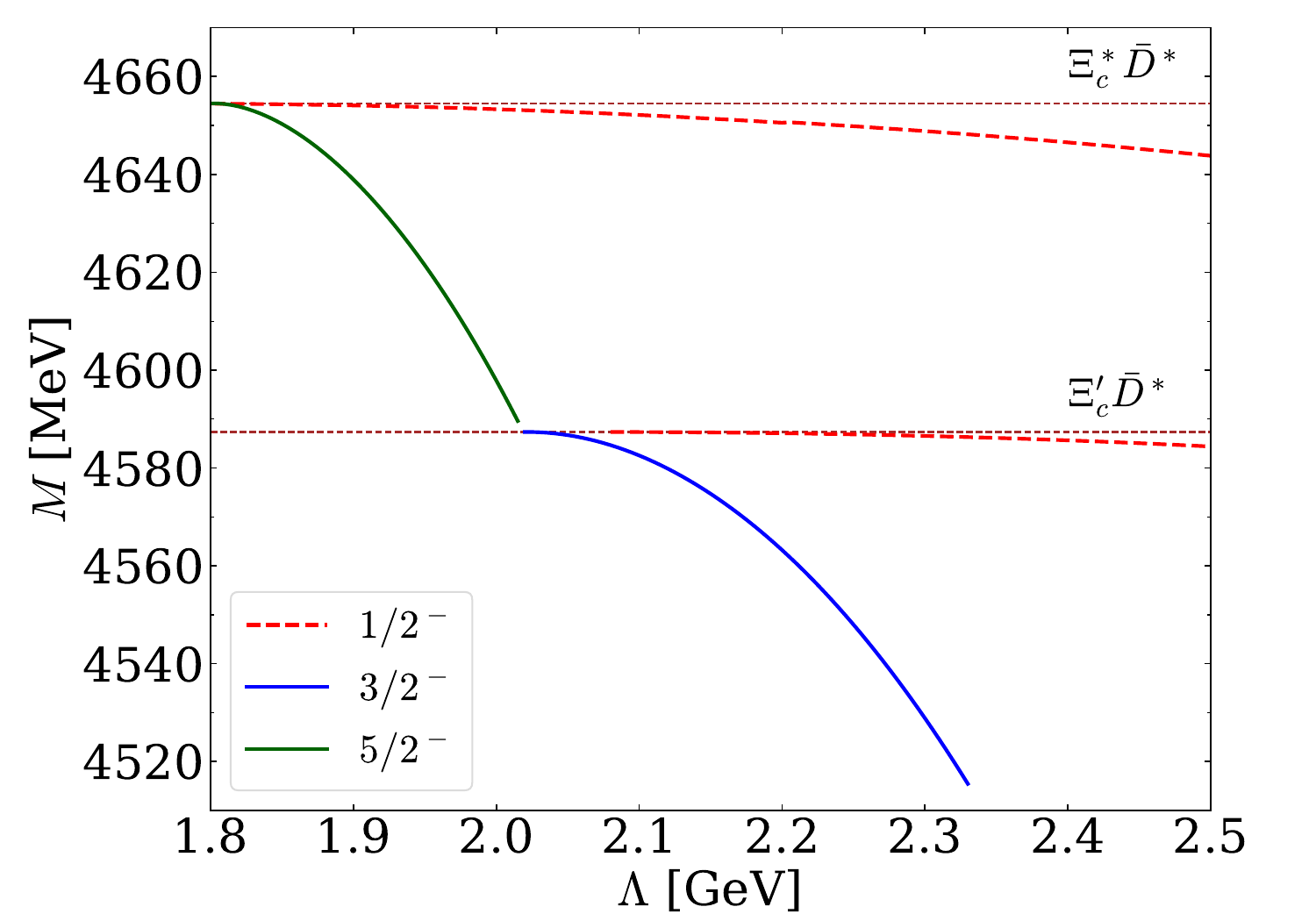}
\caption{Plots analogous to Fig.~\ref{E(I=0)} are shown for the $I=1$ system.}
\label{E(I=1)}
\end{figure}

Now, we are ready to discuss the bound states or resonances in $\Xi_c^{(\prime,*)}\bar D^{(*)}$ system with this approach. As the cutoff value $\Lambda$ varies, the mass spectra of the possible bound states in $I=0$ and $I=1$ systems are shown in Figs.~\ref{E(I=0)} and ~\ref{E(I=1)}, respectively. In each plot, the results for the case with $a=0$ and $a=1$ are presented with solid and dashed lines, respectively.  

In the $I=0$ system, ten bound states are found within the range of $\Lambda=1-2.5$ GeV. Among them, the $1/2^-(\Xi_c\bar D,\Xi_c'\bar D,\Xi_c\bar D^*,\Xi_c^*\bar D)$ and $3/2^-(\Xi_c\bar D^*)$ states are independent of the parameter $a$, and their binding energies (which equal $M-W$, where $W$ is the threshold mass of the channel) are roughly the same as the cutoff varies, as can be deduced from their potentials in Fig.\ref{V(r)_I=0}. The mass differences originate from  the reduced masses of the relevant channels. The remaining five states show significant $a$-parameter sensitivity. The $1/2^-(\Xi_c'\bar D^*,\Xi_c^*\bar D^*)$ and $3/2^-(\Xi_c^*\bar D^*)$ states can form bound states with relatively small cutoff($\Lambda\sim 1$GeV) at $a=0$ and it is difficult for them to produce bound states at $a=1$. The $1/2^-(\Xi_c^*\bar{D}^*)$ state becomes unbound at $a=1$ due to potential sign inversion. In other words, an increase in the parameter $a$ leads to a reduction in attractive potentials for this system, which eventually turns into repulsion. Conversely, the $3/2^-(\Xi_c'\bar D^*)$ state and $5/2^-(\Xi_c^*\bar D^*)$ state become deeper bound states for $a=1$, compared to situation when $a=0$.   

\begin{figure}[htbp]
	\centering 
	\includegraphics[width=0.45\textwidth]{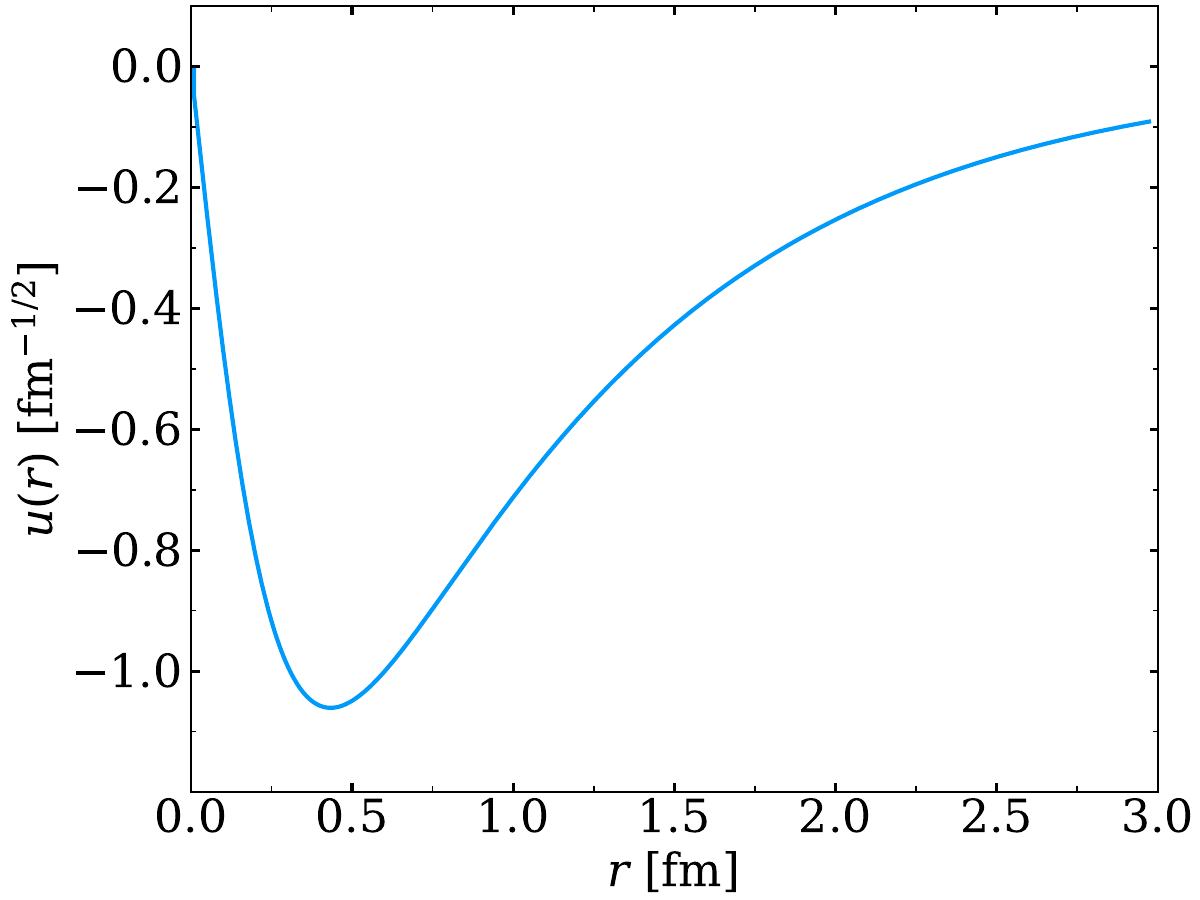}
	\caption{Bound state wave function for the $(1/2^-,3/2^-) \Xi_{c}\bar{D}^{*}$ system with $I=0$ when the cutoff $\Lambda$=2.12 GeV.}
	\label{wavefunc}
	\end{figure}

In the pioneering observation of hidden-charm pentaquarks with strangeness, the LHCb Collaboration reported $P_{cs}(4459)$ lying about $19$ MeV below the $\Xi_c\bar{D}^*$ threshold, suggesting  an interpretation as a molecular state with $J^P=1/2^-$ or $3/2^-$ quantum numbers~\cite{LHCb:2020jpq,Wu:2010vk,Feijoo:2015kts,Lu:2016roh,Chen:2016ryt,Xiao:2019gjd,Wang:2023ael}. Our analysis in Fig.~\ref{E(I=0)} demonstrates that both $1/2^-$ and $3/2^-$ states below the $\Xi_c\bar{D}^*$ threshold can reproduce the $P_{cs}(4459)$ mass when using a cutoff parameter $\Lambda=2.12$ GeV. The wave function of this bound state is shown in Fig.~\ref{wavefunc}. 
Subsequent analyses from the Belle and Belle II Collaborations revealed another state whose mass is shifted by about $13$ MeV relative to $P_{cs}(4459)$, and lies $6$ MeV below the $\Xi_c\bar{D}^*$ threshold~\cite{Belle:2025pey}. While our single-channel calculation predicts degenerate masses for the $1/2^-$ and $3/2^-$ states below the $\Xi_c\bar{D}^*$ threshold, we resolve this degeneracy through coupled channel dynamics in Sec.~\ref{Pcs4459}. Furthermore, the state with $1/2^-$ below the $\Xi_c\bar{D}$ threshold forms a bound state at $\Lambda=1.5$ GeV, potentially corresponding to the LHCb Collaboration's observation of the $P_{cs}(4338)$ state~\cite{LHCb:2022ogu}.

For the isovector systems, as shown in Fig. \ref{E(I=1)}, the OBE model reveals distinct binding patterns depending on the $\delta(\bm{r})$ term treatment. When parameter $a$ is set to $0$, we find two bound states: $3/2^-(\Xi_c'\bar{D}^*)$ and $5/2^-(\Xi_c^*\bar{D}^*)$, requiring cutoff values $\Lambda=2.0$-$2.4$ GeV and $\Lambda=1.8$-$2.0$ GeV respectively. Conversely, with the parameter $a=1$, other two $1/2^-$ states emerge: $(\Xi_c'\bar{D}^*)$ at $\Lambda=2.1$-$3.8$ GeV and $(\Xi_c^*\bar{D}^*)$ at $\Lambda=1.75$-$3.5$ GeV. The significantly higher cutoff requirements for $I=1$ systems compared to $I=0$ counterparts are due to the reduced attraction in isovector systems. This reduced attraction arises from destructive interference between isovector ($\rho$, $\pi$) and isoscalar ($\omega$, $\eta$) meson exchanges, as indicated in Fig.~\ref{V(r)_I=1}. The former two states at $a=1$ and latter two states at $a=0$ can not form bound states because their potentials become repulsive at these values of the parameter $a$. The OBE potential for the $3/2^-(\Xi_c^*\bar D^*)$ system is attractive at $a=1$, and it can form a bound state as the cutoff increases up to $4$ GeV. Given the large cutoff requirements for single-channel isovector bound states within  cases of both $a=0$ and $1$, it is expected that coupled channel dynamics likely provides rich insights for these systems, as discussed in Ref.~\cite{Yang:2022ezl}.     

To better understand the OBE dynamics in the formation of hadronic molecules, we calculate the root-mean-square (RMS) radii and potential expectation value (PEV) for the discussed bound states by varying the cutoff $\Lambda$. The RMS ($\sqrt{\langle r\rangle^2}$) and PEV ($\langle V\rangle$) for the bound state wave function $u(r)$ are calculated as
\begin{align}
	\langle r\rangle^2&=\int u(r)^\dagger r^2 u(r)dr,\\
	\langle V\rangle&=\int u(r)^\dagger V(r) u(r)dr.
\end{align} 

Tables~\ref{tab:rms_I=0} and \ref{tab:rms1_I=0} show the binding energy, RMS radius and PEV for the bound states in $\Xi_c^{(\prime,*)}\bar{D}^{(*)}$ systems with isospin $I=0$. As discussed above, the five states in Table~\ref{tab:rms_I=0} are  independent of the parameter $a$ and pseudoscalar meson exchange is forbidden in these systems. With the same cutoff range from $1.90$ to $2.12$ GeV, their RMS are around $1 \text{ fm}$ which is a typical hadronic molecular scale. The PEV for $\rho$ meson exchange provides the dominant contribution to the formation of these hadronic molecules. 

The other five bound states in Table~\ref{tab:rms1_I=0} reveal distinct patterns, since they depend on the parameter $a$. Specifically, for $a = 0$, taking the $1/2^-(\Xi_c'\bar{D}^*)$ state as an example, the binding energy $E$ is only $-0.1$ MeV at $\Lambda = 0.9$ GeV, with a relatively large RMS radius of $1.7$ fm, indicating a weak and loose molecular configuration. As $\Lambda$ increases to $1.1$ GeV, the binding energy deepens significantly to $-36.4$ MeV, and the RMS radius shrinks to $0.7$ fm. This compact configuration is mainly driven by the dominant $\pi$ meson contributions, with $\langle V_\pi\rangle=-85.3$ MeV at $a=0$. However, in $a=1$ case, dominant contributions come from the $\sigma$ meson exchange and potentials for $(\pi,\eta,\rho)$ mesons show sign inversion and become repulsive. In addition, the binding energies, RMS distances, and PEV for the bound states with isospin $I = 1$ are summarized in Table V.

\begin{table*}[ht!]
\centering
\begin{ruledtabular}
	\caption{Binding energies ($E$), RMS, and PEV for $a$-independent bound states in $\Xi_c^{(\prime,*)}\bar{D}^{(*)}$ systems with isospin $I=0$. The PEVs for each meson exchange OBE are given in the last six columns in units of $\text{MeV}$. The entries with ``$\cdots$'' indicate that such meson exchanges are forbidden.   }
\label{tab:rms_I=0}
\begin{tabular}{lcccccccccc}
State & $\Lambda$ (GeV)& $E$ (MeV)& $\sqrt{\braket{r^2}}$ (fm)& $\braket{\sigma}$ & $\braket{\pi}$ & $\braket{\eta}$ & $\braket{\rho}$ & $\braket{\omega}$ & $\braket{V_{\text{total}}}$ \\
\hline
$1/2^- ~ \Xi_{c}\bar{D}$ & 1.90 & -8.3 & 1.2 & -18.7 &$\cdots$&$\cdots$& -52.2 & 16.9 & -54.0 \\ 
& 2.00 & -11.8 & 1.1 & -22.8 &$\cdots$&$\cdots$& -65.2 & 21.1 & -66.8 \\ 
& 2.12 & -16.4 & 1.0 & -29.0 &$\cdots$&$\cdots$& -85.6 & 27.8 & -86.8\\
\hline
$1/2^- ~ \Xi_{c}^{\prime}\bar{D}$ & 1.90 & -9.2 & 1.1 & -19.6 &$\cdots$&$\cdots$& -54.7 & 17.7 & -56.6 \\ 
& 2.00 & -12.9 & 1.0 & -23.8 &$\cdots$&$\cdots$& -68.1 & 22.1 & -69.8 \\
& 2.12 & -17.7 & 0.9 & -30.1 &$\cdots$&$\cdots$& -89.1 & 29.0 & -90.3 \\
\hline
$(1/2^-,3/2^-) ~ \Xi_{c}\bar{D}^{*}$ & 1.90 & -10.3 & 1.1 & -20.7 &$\cdots$&$\cdots$& -57.8 & 18.7 & -59.8 \\ 
& 2.00 & -14.2 & 1.0 & -24.9 &$\cdots$&$\cdots$& -71.7 & 23.3 & -73.4 \\
& 2.12 & -19.4 & 0.9 & -31.5 &$\cdots$&$\cdots$& -93.4 & 30.4 & -94.5 \\
\hline
$3/2^- ~ \Xi_{c}^{*}\bar{D}$ & 1.90 & -9.7 & 1.1 & -20.1 &$\cdots$&$\cdots$& -56.2 & 18.2 & -58.1 \\ 
& 2.00 & -13.5 & 1.0 & -24.3 &$\cdots$&$\cdots$& -69.8 & 22.6 & -71.5 \\ 
& 2.12 & -18.5 & 0.9 & -30.8 &$\cdots$&$\cdots$& -91.2 & 29.6 & -92.3 \\
\end{tabular}
\end{ruledtabular}
\end{table*}

\begin{table*}[ht!]
	\begin{ruledtabular}
\caption{Similar to Table~\ref{tab:rms_I=0} for $a$-dependent bound states in $\Xi_c^{(\prime,*)}\bar{D}^{(*)}$ systems with isospin $I=0$}
\label{tab:rms1_I=0}
\begin{tabular}{lcccccccccc}
	State & $\Lambda$ (GeV)& $E$ (MeV)& $\sqrt{\braket{r^2}}$ (fm)& $\braket{\sigma}$ & $\braket{\pi}$ & $\braket{\eta}$ & $\braket{\rho}$ & $\braket{\omega}$ & $\braket{V_{\text{total}}}$ \\
\hline
$1/2^- ~ \Xi_{c}^{\prime}\bar{D}^{*} ~ a=0$ & 0.90 & -0.1 & 1.7 & -2.0 & -10.7 & -0.2 & -2.2 & 0.6 & -14.5 \\
& 1.00 & -9.1 & 1.1 & -7.1 & -38.1 & -1.3 & -17.0 & 5.1 & -58.4 \\ 
& 1.10 & -36.4 & 0.7 & -15.3 & -85.3 & -3.7 & -58.9 & 18.3 & -144.9 \\
\hline
 $1/2^- ~ \Xi_{c}^{\prime}\bar{D}^{*} ~ a=1$ & 1.70 & -2.9 & 1.4 & -10.9 & 3.1 & 1.4 & 7.0 & -2.6 & -2.0 \\ 
& 1.80 & -11.7 & 1.1 & -18.2 & 4.4 & 2.4 & 12.2 & -4.5 & -3.8 \\ 
& 1.90 & -26.8 & 0.8 & -26.1 & 5.5 & 3.4 & 18.2 & -6.8 & -5.8 \\ 
\hline
$3/2^- ~ \Xi_{c}^{\prime}\bar{D}^{*} ~ a=0$ & 1.50 & -8.7 & 1.3 & -8.8 & 9.5 & 0.3 & -17.0 & 5.4 & -10.6 \\
& 1.60 & -18.0 & 1.1 & -13.0 & 14.7 & 0.6 & -25.2 & 8.1 & -14.9 \\ 
& 1.70 & -31.5 & 0.9 & -17.8 & 21.0 & 0.9 & -34.1 & 11.0 & -19.0 \\ 
\hline
$3/2^- ~ \Xi_{c}^{\prime}\bar{D}^{*} ~ a=1$ & 1.10 & -1.0 & 1.7 & -3.3 &  -0.9 & -0.2 & -8.4 & 2.6 &  -10.3 \\ 
& 1.20 & -7.0 & 1.3 & -7.3 & -1.5 & -0.5 & -22.3 & 7.1 & -24.5 \\ 
& 1.30 & -19.0 & 1.0 & -12.7 & -2.0 & -0.9 & -44.3 & 14.2 & -45.7 \\ 
\hline
$1/2^- ~ \Xi_{c}^{*}\bar{D}^{*} ~ a=0$ & 0.90 & -4.2 & 1.3 & -3.7 & -28.7 & -0.7 & -5.0 & 1.4 & -36.7 \\
& 1.00 & -24.9 & 0.8 & -10.1 & -75.3 & -2.7 & -30.1 & 9.1 & -109.2 \\ 
& 1.10 & -71.0 & 0.6 & -19.1 & -146.4 & -6.6 & -92.7 & 28.8 & -236.0\\
\hline
$3/2^- ~ \Xi_{c}^{*}\bar{D}^{*} ~ a=0$ & 1.10 & -3.0 & 1.4 & -6.0 & -12.5 & -0.5 & -13.8 & 4.2 & -28.6 \\ 
& 1.20 & -13.7 & 1.0 & -13.0 & -28.9 & -1.3 & -39.3 & 12.4 & -70.1 \\ 
& 1.30 & -33.9 & 0.7 & -21.9 & -52.9 & -2.8 & -83.3 & 26.5 & -134.4 \\
\hline
$3/2^- ~ \Xi_{c}^{*}\bar{D}^{*} ~ a=1$ & 2.00 & -1.5 & 1.5 & -10.8 & 1.4 & 0.7 & -11.2 & 3.4 & -16.4 \\
& 2.12 & -2.7 & 1.4 & -13.5 & 1.6 & 0.9 & -14.4 & 4.4 & -21.0 \\
& 2.20 & -3.6 & 1.4 & -15.5 & 1.8 & 1.0 & -16.7 & 5.2 & -24.3 \\ 
\hline
$5/2^- ~ \Xi_{c}^{*}\bar{D}^{*} ~ a=0$ & 2.00 & -2.7 & 1.6 & -5.8 & 4.3 & 0.0 & -16.6 & 5.4 & -12.8 \\
& 2.12 & -4.1 & 1.5 & -7.0 & 5.0 & 0.0 & -20.8 & 6.7 & -16.0 \\ 
& 2.20 & -5.2 & 1.4 & -7.9 & 5.6 & 0.0 & -23.9 & 7.8 & -18.4 \\
\hline
$5/2^- ~ \Xi_{c}^{*}\bar{D}^{*} ~ a=1$ & 1.10 & -1.3 & 1.7 & -3.5 & -1.4 & -0.4 & -10.6 & 3.3 & -12.6 \\
& 1.20 & -6.3 & 1.3 & -7.1 & -2.1 & -0.7 & -26.0 & 8.3 & -27.8 \\ 
& 1.30 & -14.9 & 1.1 & -12.0 & -2.9 & -1.2 & -49.7 & 15.9 & -49.8 \\ 
\end{tabular}
\end{ruledtabular}
\end{table*}

\begin{table*}[ht]
	\begin{ruledtabular}
\caption{Similar to Table~\ref{tab:rms_I=0} for the bound states in $\Xi_c^{(\prime,*)}\bar{D}^{(*)}$ systems with isospin $I=1$.}
\label{tab:coupling_constants}
\begin{tabular}{lcccccccccc}
State & $\Lambda$ (GeV)& $E$ (MeV)& $\sqrt{\braket{r^2}}$ (fm)& $\braket{\sigma}$ & $\braket{\pi}$ & $\braket{\eta}$ & $\braket{\rho}$ & $\braket{\omega}$ & $V_{\text{total}}$ \\
\hline
$1/2^- ~ \Xi_{c}^{\prime}\bar{D}^{*} ~ a=1$ & 2.20 & -0.3 & 1.7 & -8.9 & -0.8 & 1.1 & -2.1 & -2.3 & -13.0 \\
& 2.35 & -1.2 & 1.6 & -12.7 & -1.0 & 1.6 & -3.1 & -3.5 & -18.7 \\
& 2.50 & -3.0 & 1.4 & -17.6 & -1.2 & 2.2 & -4.4 & -4.9 & -25.9 \\
\hline
$3/2^- ~ \Xi_{c}^{\prime}\bar{D}^{*} ~ a=0$ & 2.05 & -0.6 & 1.5 & -19.9 & -16.2 & 3.6 & -1.7 & -1.8 & -36.0 \\
& 2.12 & -7.5 & 1.0 & -39.3 & -36.2 & 8.5 & -11.2 & -11.3 & -89.5 \\
& 2.20 & -24.1 & 0.7 & -60.3 & -62.7 & 15.2 & -31.3 & -31.2 & -170.3 \\ 
\hline
$1/2^- ~ \Xi_{c}^{*}\bar{D}^{*} ~ a=1$ & 2.00 & -1.2 & 1.6 & -10.9 & -1.2 & 1.8 & -5.7 & -6.0 & -22.0 \\
& 2.12 & -2.5 & 1.4 & -14.5 & -1.4 & 2.3 & -7.8 & -8.2 & -29.7 \\
& 2.20 & -3.8 & 1.3 & -17.3 & -1.6 & 2.8 & -9.5 & -10.0 & -35.5 \\
\hline
$5/2^- ~ \Xi_{c}^{*}\bar{D}^{*} ~ a=0$ & 1.85 & -4.2 & 1.1 & -34.7 & -49.7 & 11.1 & -29.8 & -29.4 & -132.4 \\
& 1.90 & -15.6 & 0.7 & -52.1 & -81.5 & 18.8 & -58.0 & -57.2 & -229.9 \\
& 1.95 & -33.1 & 0.5 & -66.2 & -111.6 & 26.4 & -90.4 & -89.3 & -331.1 \\ 
\end{tabular}
\end{ruledtabular}
\end{table*}

\subsection{$\Xi_c'\bar D-\Xi_c\bar D^*$ coupled channel}
\label{Pcs4459}
In this subsection, we discuss the mass gap between $J^P=1/2^-$ and $3/2^-$ isoscalar states below the $\Xi_{c}\bar{D}^{*}$ threshold. Based on our previous single channel calculation, these two states have equal mass and correspond to the $P_{cs}(4459)$ pentaquark in the earlier observation by the LHCb Collaboration~\cite{LHCb:2020jpq}, which was reported to have mass of $4458.8 \pm 2.9^{+4.7}_{-1.1}$ MeV and width of $17.3 \pm 6.5^{+8.0}_{-5.7}$ MeV. However, in the recent observation of the Belle and Belle II Collaborations~\cite{Belle:2025pey}, its mass and width were measured to be $4471.7 \pm 4.8 \pm 0.6$ MeV and $21.9 \pm 13.1 \pm 2.7$ MeV, respectively. Interestingly, the masses of these states lie 19 and 6 MeV below the $\Xi_{c}\bar{D}^{*}$ threshold. We suggest that the two pentaquarks, $P_{cs}(4459)$ and $P_{cs}(4472)$ in our notation, could correspond to the two isoscalar states below the $\Xi_{c}\bar{D}^{*}$ threshold. 

\begin{figure}[ht!]
	\centering
	\subfigure[$\Lambda=2070$ MeV]{\label{V340}\includegraphics[width=0.48\textwidth]{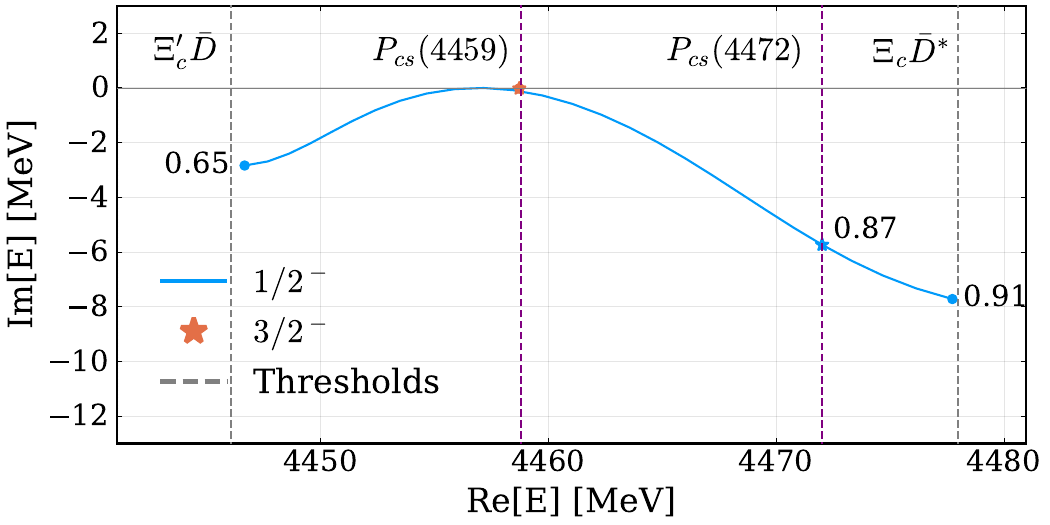}} 
	\subfigure[$\Lambda=1720$ MeV]{\label{V3402}\includegraphics[width=0.48\textwidth]{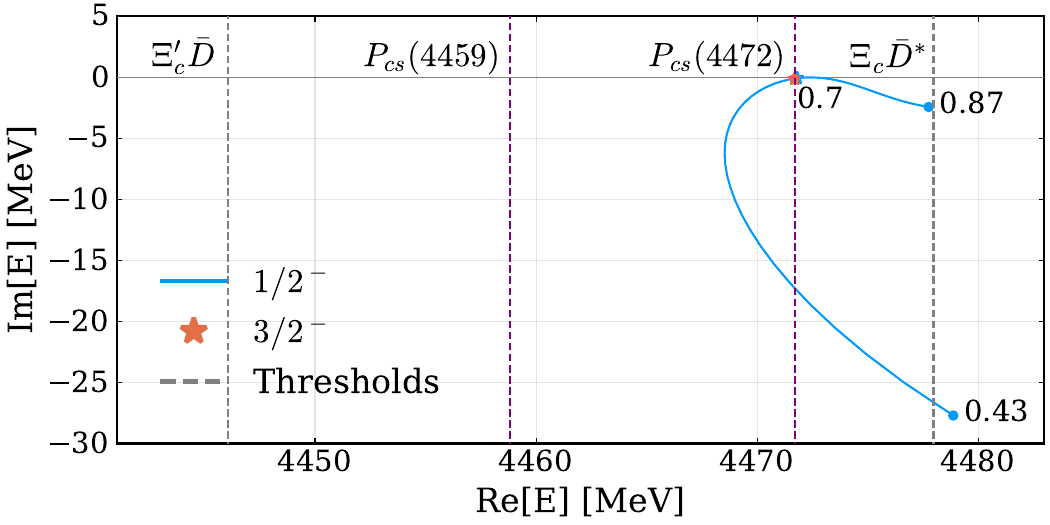}} 
	\caption{Trajectories of the poles in the $\Xi_c'\bar D-\Xi_c\bar D^*$ coupled channel system with $I=0$ by varying the parameter $a$ and fixing the cutoff. The numbered labels on the blue curves represent the value of the parameter $a$, and $a$ is smoothly varied between these values. The effect of the parameter $a$ on $3/2^-$ state is negligible thus its pole position is labeled with orange star.   }
	\label{fig:pole-troject}
\end{figure}

To explain the mass splitting between these two isoscalar states, we consider the coupled channel dynamics of two close channels $\Xi_c'\bar D$ and $\Xi_c\bar D^*$ with a threshold difference of about 32 MeV. Then we investigate the possibility of simultaneously reproducing the masses of $P_{cs}(4459)$ and $P_{cs}(4472)$ by using the same phenomenological parameters $a$ and $\Lambda$. In this coupled channel calculation, we find that the pole position of the $3/2^-$ state changes negligibly as we vary the parameter $a$. Therefore, we adjust the parameter $\Lambda$ to reproduce the mass of either $P_{cs}(4459)$ or $P_{cs}(4472)$ for the $3/2^-$ state. Specifically, we find that the $3/2^-$ state reproduces the $P_{cs}(4459)$ mass when $\Lambda = 2070$ MeV and the $P_{cs}(4472)$ mass when $\Lambda = 1720$ MeV. With $\Lambda$ fixed at these values, we show the pole trajectories for the $3/2^-$ and $1/2^-$ states as we vary the parameter $a$ in Fig. \ref{fig:pole-troject}.

In Fig.~\ref{fig:pole-troject}, the pole positions are calculated on the $(-+)$ Riemann sheet using the method described in Sec. \ref{bound energy}, searching the energy region between the $\Xi_c'\bar D$ and $\Xi_c\bar D^*$ thresholds. These poles correspond to physical resonances which can generate a peak-like structure in the scattering amplitude~\cite{Garzon:2012np}. In Fig.~\ref{V340}, we can see that the $3/2^-$ and $1/2^-$ states can reproduce the masses of the $P_{cs}(4459)$ and $P_{cs}(4472)$ pentaquarks with $\Lambda=2070$ MeV and $a=0.87$. However, as shown in Fig.~\ref{V3402}, if we set $P_{cs}(4472)$ as the $3/2^-$ state by adjusting $\Lambda$ to be $1720$ MeV, the $1/2^-$ state is well above the experimental mass of $P_{cs}(4459)$. Our calculation suggest that the former scenario is valid for simultaneously reproducing the $P_{cs}(4459)$ and $P_{cs}(4472)$ states,  their $J^P$ quantum numbers being $3/2^-$ and $1/2^-$, respectively. This spin-parity assignment is consistent with Ref.~\cite{Du:2021bgb}, in which the $J^P$ of the lower and higher states below the $\Xi_c\bar D^*$ threshold are identified as $3/2^-$ and $1/2^-$. 

Since the masses of both $P_{cs}(4459)$ and $P_{cs}(4472)$ pentaquarks are reproduced in this work with the values of $a=0.87$ and $\Lambda=2070$ MeV, other molecular states in $\Xi_{c}^{(\prime, *)}\bar{D}^{(*)}$ systems are further calculated as a prediction of the OBE model, which is presented in Table \ref{2070}. In the isoscalar system, the ten states can form the bound states. Among them, the three states with $ 1/2^-(\Xi_{c}'\bar{D}^{*})$, $3/2^-(\Xi_{c}'\bar{D}^{*})$ and $5/2^-(\Xi_{c}^{*}\bar{D}^{*})$ are deeply bound with binding energies smaller than $\sim -90$ MeV, hence they may be unphysical. The state with $1/2^-(\Xi_{c}\bar{D})$ has a binding energy of -14.4 MeV, which is consistent with the momentum space calculation in Ref.~\cite{Zhu:2021lhd}. We expect that a more comprehensive result can be captured in the calculation considering the full coupled channel dynamics of these six channels, in order to better characterize decay patterns and resolve the current degeneracy in bound state predictions. 
\begin{table}[ht]
	\caption{Prediction of the mass spectra of the molecular states in $\Xi_{c}^{(\prime, *)}\bar{D}^{(*)}$ systems with $I=0$. The thresholds ($W$), binding energies ($E$), and masses ($M$) of the molecular states are given in MeV. The masses of the two molecular states corresponding to the experimental masses of the Pcs(4459) and Pcs(4472) pentaquarks are indicated in boldface.}
	\label{2070}
	\begin{ruledtabular}
	\begin{tabular}{lccc}  
	$J^P$ (channel)                                 & $W$       & $E$      & $M$ \\ \hline
	$1/2^-(\Xi_{c}\bar{D})$                         & 4336.7    &-14.4          & 4322.2\\ 
	$1/2^-(\Xi_{c}'\bar{D})$                        & 4446.0    &-15.6          & 4430.4\\ 
	$1/2^-(\Xi_{c}\bar{D}^{*}-\Xi_{c}'\bar{D})$     & 4478.0    &-6.3           & \textbf{4471.7}\\ 
	$3/2^-(\Xi_{c}\bar{D}^{*}-\Xi_{c}'\bar{D})$     & 4478.0    &-19.2          & \textbf{4458.8}\\ 
	$3/2^-(\Xi_{c}^{*}\bar{D})$                     & 4513.2    &-16.3          & 4496.8\\ 
	$1/2^-(\Xi_{c}'\bar{D}^{*})$                    & 4587.4    &-111.9         & 4475.5\\ 
	$3/2^-(\Xi_{c}'\bar{D}^{*})$                    & 4587.4    &-204.9         & 4382.5 \\ 
	$1/2^-(\Xi_{c}^{*}\bar{D}^{*})$                 & 4654.5    &-13.4          & 4641.1\\ 
	$3/2^-(\Xi_{c}^{*}\bar{D}^{*})$                 & 4654.5    &-12.4          & 4642.1\\ 
	$5/2^-(\Xi_{c}^{*}\bar{D}^{*})$                 & 4654.5    &-91.0          & 4563.5\\ 
	\end{tabular}
	\end{ruledtabular}
	\end{table}
\subsection{Role of the $\Lambda\eta_c$ and $\Lambda J/\psi$ channels }    

The $\Lambda\eta_c$ and $\Lambda J/\psi$ channels represent closed flavor channels that all predicted states can decay into via $D^{(*)}$ meson exchange. To investigate the effects of these decay channels on our results, we include them in our analysis. The details of the corresponding potentials are provided in Appendix~\ref{app:decay}. 

With these potentials, we extract the scattering $S$-matrix by solving the coupled channel Schr\"odinger equation, and the poles are sought near the thresholds of $\Xi_c^{(',*)}\bar D^{(*)}$ channels. The bound states in Table~\ref{2070} now become resonances, and their pole positions acquire non-zero imaginary parts due to the inclusion of lower-lying channels. In our calculation, we find that the inclusion of the $\Lambda\eta_c$ and $\Lambda J/\psi$ channels has a marginal effect on the masses of the states in Table~\ref{2070} as the parameter $a$ varies from 0 to 1. Such marginal effects are also expected because the masses of these predicted states are significantly above the thresholds of the $\Lambda\eta_c$ and $\Lambda J/\psi$ channels, and the coupling constants $G_1$, $G_2$, and $G_3$ given in Appendix~\ref{app:decay} are relatively small. 

To better visualize the role of these two channels, we introduce a factor $x$ that multiplies each of the coupling constants $G_1$, $G_2$, and $G_3$, and recalculate the poles in Table~\ref{2070} while including the $\Lambda\eta_c$ and $\Lambda J/\psi$ channels. The results for $x=1$, $1.5$, and $2$ are listed in Table~\ref{tab:decay}. Among these poles, the greatest effect is observed for the first pole $1/2^-(\Xi_{c}\bar{D})$, whose imaginary part changes from $-0.17$ to $-3.09$ as $x$ increases from $1$ to $2$. The third pole $1/2^-(\Xi_{c}\bar{D}^{*})$ already has an imaginary part corresponding to the resonance in the $\Xi_c'\bar D-\Xi_{c}\bar{D}^{*}$ coupled channel (as shown in Fig.~\ref{fig:pole-troject}), and the inclusion of the $\Lambda\eta_c$ and $\Lambda J/\psi$ channels has only a tiny effect on its pole position. The final pole $5/2^-(\Xi_{c}^{*}\bar{D}^{*})$ remains unchanged to at least two floating-point precision as $x$ increases, due to the large energy gap between its mass and the $\Lambda\eta_c$ and $\Lambda J/\psi$ channel thresholds.

\begin{table}[ht]
	\caption{Pole positions of the molecular states after including the coupled channel dynamics of $\Lambda\eta_c$ and $\Lambda J/\psi$ channels, compared with the results in Table~\ref{2070}. The nearby threshold channels are indicated in parentheses. All pole positions are given in units of MeV. }
	\label{tab:decay}
	\begin{ruledtabular}
	\begin{tabular}{lccc}  
	$J^P$(channel)                                 & $x=1$       & $x=1.5$      & $x=2$ \\ \hline
	$1/2^-(\Xi_{c}\bar{D})$                         & $4322.05$-$i 0.17$    & $4321.13$-$i 0.87$         & $4318.48$-$i 3.09$ \\ 
	$1/2^-(\Xi_{c}'\bar{D})$                        & $4430.43$-$i 0.00$    & $4430.40$-$i 0.01$        & $4430.33$-$i 0.04$ \\ 
	$1/2^-(\Xi_{c}\bar{D}^{*})$     & $4471.83$-$i 5.81$     & $4471.69$-$i 5.92 $      & $4471.29$-$i 6.23$ \\ 
	$3/2^-(\Xi_{c}\bar{D}^{*})$     & $4458.83$-$i 0.01$    & $4458.79$-$i 0.01$        & $4458.70$-$i 0.02 $\\ 
	$3/2^-(\Xi_{c}^{*}\bar{D})$                     & $4496.88$-$i 0.00$    & $4496.88$-$i 0.00$        & $4496.86$-$i 0.01$ \\ 
	$1/2^-(\Xi_{c}'\bar{D}^{*})$                    & $4475.44$-$i 0.02$    & $4475.33$-$i 0.09$        & $4475.03$-$i 0.28$ \\ 
	$3/2^-(\Xi_{c}'\bar{D}^{*})$                    & $4290.88$-$i 0.00$    & $4290.85$-$i 0.00$         & $4290.78$-$i 0.01 $ \\ 
	$1/2^-(\Xi_{c}^{*}\bar{D}^{*})$                 & $4642.37$-$i 0.00$    & $4642.37$-$i 0.01 $          & $4642.38$-$i 0.04$  \\ 
	$3/2^-(\Xi_{c}^{*}\bar{D}^{*})$                 & $4642.23$-$i 0.01 $   & $4642.22$-$i 0.03 $            & $4642.17$-$i 0.08 $\\ 
	$5/2^-(\Xi_{c}^{*}\bar{D}^{*})$                 & $4563.50$-$i 0.00 $   & $4563.50$-$i 0.00$             & $4563.50$-$i 0.00$ \\ 
	\end{tabular}
	\end{ruledtabular}
	\end{table}

For the $P_{cs}(4459)$ and $P_{cs}(4472)$ pentaquark candidates, the two states with $1/2^-$ and $3/2^-$ spin-parity quantum numbers in the $\Xi_c'\bar D-\Xi_{c}\bar{D}^{*}$ coupled channel system may exhibit different behaviors when decaying to $\Lambda\eta_c$ and $\Lambda J/\psi$ channels. Therefore, we calculate the scattering $T$ matrix for the coupled channel system $\Lambda\eta_c-\Lambda J/\psi-\Xi_c'\bar D-\Xi_{c}\bar{D}^{*}$. Following Refs.~\cite{Doring:2009yv,Garzon:2012np}, the $T(E)$ matrix can be related to $S(E)$ in Eq.~\eqref{eq:asym-wave} as,
\begin{align}
	S_{jk}(E)=1+i\sqrt{2\rho_j}T_{jk}(E)\sqrt{2\rho_k},
\end{align} 
where $j$ and $k$ are channel indices. In non-relativistic approximation, two body phase space factor $\rho_j$ for channel $j$ can be written as function of channel momentum $q_j(E)$ as
\begin{align}
	\rho_j=\frac{q_j(E)}{8\pi E}.
\end{align} 

\begin{figure}[htbp]
\centering 
\includegraphics[width=0.5\textwidth]{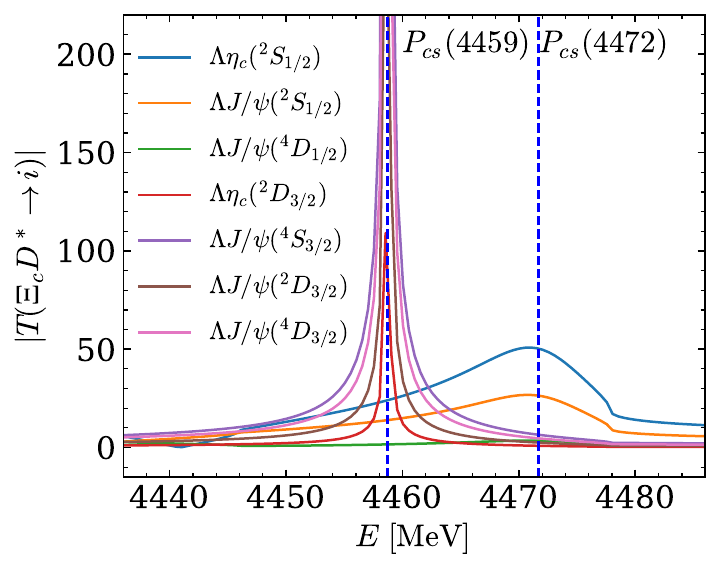}
\caption{$T$ matrix for the scattering of the $\Xi_{c}\bar{D}^{*}$ channel to $\Lambda\eta_c$ and $\Lambda J/\psi$ channels with $x=1$.}
\label{fig:TE}
\end{figure}

Fig.~\ref{fig:TE} shows the energy-dependent scattering $T$ matrix elements for transitions from the $S$-wave $\Xi_{c}\bar{D}^{*}$ channel to $\Lambda\eta_c$ and $\Lambda J/\psi$ channels, where each line represents $T(\Xi_c\bar D^*\to i)$ with $i$ denoting the partial wave channels in $\Lambda\eta_c$ and $\Lambda J/\psi$. The results are obtained using the same model parameters as in Table~\ref{tab:decay} with $x=1$. Since the $S$-wave $\Xi_{c}\bar{D}^{*}$ channel dominates both $1/2^-$ and $3/2^-$ states, the peak-like behavior of the $T$ matrix elements for transitions from the $S$-wave $\Xi_{c}\bar{D}^{*}$ channel to other channels is relatively more pronounced. It is observed that the energy corresponding to the peak of the $T$ matrix in the coupled channel system with $1/2^-$ corresponds to the $P_{cs}(4472)$ mass, while that with $3/2^-$ corresponds to the $P_{cs}(4459)$ mass. In the $\Lambda\eta_c-\Lambda J/\psi-\Xi_c'\bar D-\Xi_{c}\bar{D}^{*}$ coupled channel system with $J^P=1/2^-$, the $\Xi_{c}\bar{D}^{*}(^2S_{1/2})\to\Lambda\eta_c(^2S_{1/2})$ transition is stronger compared to others, and the production rate of the $\Lambda\eta_c(^2S_{1/2})$ channel is more significant, making it easier to detect the $P_{cs}(4472)$ pentaquark in this channel than in others. For the $J^P=3/2^-$ system of these coupled channels, the peak corresponding to $P_{cs}(4459)$ is much narrower, and the production rate of the $\Lambda J/\psi(^4S_{3/2})$ channel is more significant than others, indicating that this channel is important for detecting the $P_{cs}(4459)$ pentaquark. We also observe that changing the value of parameter $x$ does not alter the patterns observed above.
\section{Summary and conclusion}~\label{sec:summary}
In this work, we have investigated the hidden-charm hadronic molecular states in $\Xi_c^{(\prime,*)}\bar D^{(*)}$ system in the context of the OBE model. The effective potentials are derived from a Lagrangian that respects both heavy quark spin symmetry and SU(3)-flavor symmetry. We analyzed possible near-threshold molecular states with quantum numbers $J^P=1/2^{-}$, $3/2^{-}$ and $5/2^{-}$ by analytically continuing the non-relativistic scattering S-matrix, which was extracted from asymptotic wave functions obtained by solving the coupled channel Schr\"odinger equation.

A key aspect of our analysis was to investigate the role of the short-range $\delta(\bm r)$ term in the OBE model. By introducing a parameter $a$ to control its contribution, we systematically studied how this term affects the bound states. Our analysis reveals ten bound states in $\Xi_c^{(\prime,*)}\bar{D}^{(*)}$ systems with isospin $I=0$ within the cutoff range $\Lambda=1-2.5$ GeV. Five states show strong dependence on the $\delta(\bm{r})$ term controlled by the parameter $a$, while the remaining five demonstrate $a$-independent binding. In the single-channel analysis, both $1/2^-$ and $3/2^-$ states below the $\Xi_c\bar{D}^*$ threshold can reproduce $P_{cs}(4459)$ pentaquark at $\Lambda=2.12$ GeV. The $1/2^-(\Xi_c\bar{D})$ state can form a bound state below the $\Xi_c\bar{D}$ threshold with relativity lower cutoff of $\Lambda=1.5$ GeV. But, our model has diffuculty reproducing the $P_{cs}(4338)$ pentaquark as $\Xi_c\bar{D}$ molecule, as it is above the threshold of $\Xi_c\bar{D}$. It is expected that the coupled channel dynamics of $\Lambda_c \bar D_s-\Xi_c\bar{D}$ coupled with $K^*$ exchange and short-range dynamics studied in this work together may provide great insights into the molecular nature of $P_{cs}(4338)$. For the isospin $I=1$, the OBE potentials for various single-channels in $\Xi_c^{(\prime,*)}\bar{D}^{(*)}$ are reduced due to destructive interference between the isoscalar and isovector meson exchanges. Thus, only two bound states separately in the marginal case of $a=0$ and $a=1$ can be formed with cutoff larger than $1.8$ GeV. Considering the coupled channel dynamics in isovector systems may reveal additional possibilities that merit further investigation.

In the coupled channel analysis of $\Xi_c'\bar{D}-\Xi_c\bar{D}^*$, we resolve the mass splitting between the LHCb Collaboration's $P_{cs}(4459)$ and the Belle Collaboration's $P_{cs}(4472)$ pentaquarks, identifying them as $3/2^-$ and $1/2^-$ isoscalar states, respectively. We use the masses of these two states to fix the two phenomenological paramters $a$ and $\Lambda$, and simultaneously predict the molecular states in the isoscalar sector. Our predictions highlight several promising molecular candidates with $I=0$: $1/2^-(\Xi_c\bar{D})$, $1/2^-(\Xi_c'\bar{D}^*)$, $3/2^-(\Xi_c^*\bar{D})$, $1/2^-(\Xi_c^*\bar{D}^*)$, and $3/2^-(\Xi_c^*\bar{D}^*)$, some of them could manifest as narrow resonances in 4.3-4.7 GeV energy region. The predicted spectrum exhibits a richness in bound and resonant states that are not yet fully probed by experiments. This feature stems from the specific form factor regularization scheme and sensitivity to short-range dynamics. A more realistic description, such as a sophisticated coupled channel treatment or the inclusion of quark substructure effects, could mitigate this excess attraction and provide a more precise phenomenology. 

Finally, we investigate the role of $\Lambda\eta_c$ and $\Lambda J/\psi$ decay channels in the phenomenology of the predicted molecular states. Our analysis shows that while these decay channels have marginal effects on the masses of the bound states, they significantly influence the imaginary parts of the pole positions, particularly for states near the decay thresholds. This analysis provides valuable insights into the detectability of the predicted molecular candidates through different production mechanisms, with the $1/2^-(\Xi_c\bar D^*)$ state showing enhanced production rates in the $\Lambda\eta_c$ channel and the $3/2^-(\Xi_c\bar D^*)$ state being more prominent in the $\Lambda J/\psi$ channel. These results offer important guidance for future experimental searches of these hidden-charm pentaquark candidates.

\begin{acknowledgments}	
	This work is supported by the Natural Science Foundation of the Xinjiang Uyghur Autonomous Region of China under Grants No. 2025D01C292 and No. 2022D01C52 and by the Doctoral Program of Tian Chi Foundation of the Xinjiang Uyghur Autonomous Region of China under grant No. 51052300506. The work of Y. R. and N. Y. is further supported by the National Natural Science Foundation of China under Grant No. 12565016.
\end{acknowledgments}

\section*{DATA AVAILABILITY}
    The data that support the findings of this article are openly available \cite{LHCb:2022ogu,ParticleDataGroup:2024cfk,Belle:2025pey}.


\appendix
\section{OBE effective potentials related to $\Xi_c^{(\prime,*)}\bar D^{(*)}$ systems}
\label{sec:OBE-pot}

Here, we list the OBE potentials for the molecular states of $\Xi_c^{(\prime,*)}\bar D^{(*)}$ systems, they can be expressed in terms of the functions in Eqs.~\eqref{eq:Yr}, ~\eqref{eq:Cr} and ~\eqref{eq:Tr}.
\begin{align}
V_{\Xi_{c}\bar{D}\rightarrow\Xi_{c}\bar{D}}&= 2l_{B}g_{s}\chi^{\dagger}_{3}\chi_{1}Y_\sigma+\frac{I_F}{4}\beta\beta_{B}g_{v}^{2}\chi^{\dagger}_{3}\chi_{1}Y_\rho\notag\\
&+\frac{1}{4}\beta\beta_{B}g_{v}^{2}\chi^{\dagger}_{3}\chi_{1}Y_\omega,
\end{align}
\begin{align}
V_{\Xi_{c}'\bar{D}\rightarrow\Xi_{c}'\bar{D}}&= -l_{s}g_{s}\chi^{\dagger}_{3}\chi_{1}Y_\sigma-\frac{I_F}{8}\beta\beta_{s}g_{v}^{2}\chi^{\dagger}_{3}\chi_{1}Y_\rho\notag\\
&-\frac{1}{8}\beta\beta_{s}g_{v}^{2}\chi^{\dagger}_{3}\chi_{1}Y_\omega,
\end{align}
\begin{align}
\label{V33}
V_{\Xi_{c}\bar{D}^{*}\rightarrow\Xi_{c}\bar{D}^{*}}&= 2 l_{B}g_{s}\chi^{\dagger}_{3}\chi_{1}\bm{\epsilon}_{4}^{*}\cdot\bm{\epsilon}_{2}Y_\sigma+\frac{I_F}{4}\beta\beta_{B}g_{v}^{2}\chi^{\dagger}_{3}\chi_{1}\bm{\epsilon}_{4}^{*}\cdot\bm{\epsilon}_{2}Y_\rho\notag\\
&+\frac{1}{4}\beta\beta_{B}g_{v}^{2}\chi^{\dagger}_{3}\chi_{1}\bm{\epsilon}_{4}^{*}\cdot\bm{\epsilon}_{2}Y_\omega,
\end{align}
\begin{align}
V_{\Xi_{c}^{*}\bar{D}\rightarrow\Xi_{c}^{*}\bar{D}}&=-l_{S}g_{S}\bm{\chi}^{\dagger}_{3}\cdot\bm{\chi}_{1}Y_\sigma-\frac{I_F}{4}\beta\beta_{S}g_{V}^{2}\bm{\chi}_{3}^{\dagger}\cdot\bm{\chi}_{1}Y_\rho\notag\\
&-\frac{1}{8}\beta\beta_{S}g_{V}^{2}\bm{\chi}_{3}^{\dagger}\cdot\bm{\chi}_{1}Y_\omega,
\end{align}
\begin{align}
V_{\Xi_{c}'\bar{D}^{*}\rightarrow\Xi_{c}'\bar{D}^{*}}&= -l_{S}g_{S}\chi^{\dagger}_{3}\chi_{1}\bm{\epsilon}_{4}^{*}\cdot\bm{\epsilon}_{2}Y_\sigma\notag\\
&-\frac{I_Fgg_{1}}{12f_{\pi}^2}\chi_{3}^{\dagger}[\bm{\sigma}\cdot(i\bm{\epsilon}_{2}\times\bm{\epsilon}_{4}^{*})C_\pi\notag\\
&+S(\bm{\sigma},i\bm{\epsilon}_{2}\times\bm{\epsilon}_{4}^{*},\hat{r})T_\pi]\chi_{1}\notag\\
&+\frac{gg_{1}}{36f_{\pi}^2}\chi_{3}^{\dagger}[\bm{\sigma}\cdot(i\bm{\epsilon}_{2}\times\bm{\epsilon}_{4}^{*})C_\eta\notag\\
&+S(\bm{\sigma},i\bm{\epsilon}_{2}\times\bm{\epsilon}_{4}^{*},\hat{r})T_\eta]\chi_{1}\notag\\
&-\frac{I_F}{8}\beta\beta_{S}g_{V}^{2}\chi_{3}^{\dagger}\chi_{1}(\bm{\epsilon}_{4}^{*}\cdot\bm{\epsilon}_{2})Y_\rho\notag\\
&-\frac{I_F}{18}\lambda\lambda_{S}g_{V}^{2}\chi_{3}^{\dagger}[ 2\bm{\sigma}\cdot(i\bm{\epsilon}_{2}\times\bm{\epsilon}_{4}^{*})C_\rho\notag\\
&-S(\bm{\sigma},i\bm{\epsilon}_{2}\times\bm{\epsilon}_{4}^{*},\hat{r})T_\rho ]\chi_{1}\notag\\
&-\frac{1}{8}\beta\beta_{S}g_{V}^{2}\chi_{3}^{\dagger}\chi_{1}(\bm{\epsilon}_{4}^{*}\cdot\bm{\epsilon}_{2})Y_\omega\notag\\
&-\frac{1}{18}\lambda\lambda_{S}g_{V}^{2}\chi_{3}^{\dagger}[ 2\bm{\sigma}\cdot(i\bm{\epsilon}_{2}\times\bm{\epsilon}_{4}^{*})C_\omega\notag\\
&-S(\bm{\sigma},i\bm{\epsilon}_{2}\times\bm{\epsilon}_{4}^{*},\hat{r})T_\omega ]\chi_{1},
\end{align}
\begin{align}
V_{\Xi_{c}^{*}\bar{D}^{*}\rightarrow\Xi_{c}^{*}\bar{D}^{*}}=& -l_{S}g_{S}(\bm{\chi}^{\dagger}_{3}\cdot\bm{\chi}_{1})(\bm{\epsilon}_{4}^{*}\cdot\bm{\epsilon}_{2})Y_\sigma\notag\\
&+\frac{I_Fgg_{1}}{8f_{\pi}^{2}}[(i\bm{\chi}^{\dagger}_{3}\times\bm{\chi}_{1})\cdot(i\bm{\epsilon}_{2}\times\bm{\epsilon}_{4}^{*})C_\pi\notag\\
&+S(i\bm{\chi}^{\dagger}_{3}\times\bm{\chi}_{1},i\bm{\epsilon}_{2}\times\bm{\epsilon}_{4}^{*},\hat{r})T_\pi ]\notag\\
&-\frac{gg_{1}}{24f_{\pi}^{2}}[(i\bm{\chi}^{\dagger}_{3}\times\bm{\chi}_{1})\cdot(i\bm{\epsilon}_{2}\times\bm{\epsilon}_{4}^{*})C_\eta\notag\\
&+S(i\bm{\chi}^{\dagger}_{3}\times\bm{\chi}_{1},i\bm{\epsilon}_{2}\times\bm{\epsilon}_{4}^{*},\hat{r})T_\eta]\notag\\
&- \frac{I_F}{8}\beta\beta_{S}g_{V}^{2}(\bm{\chi}^{\dagger}_{3}\cdot\bm{\chi}_{1})(\bm{\epsilon}_{4}^{*}\cdot\bm{\epsilon}_{2})Y_\rho\notag\\
&+\frac{I_F}{12}\lambda\lambda_{S}g_{V}^{2}[ 2(i\bm{\chi}^{\dagger}_{3}\times\bm{\chi}_{1})\cdot(i\bm{\epsilon}_{2}\times\bm{\epsilon}_{4}^{*})C_\rho\notag\\
&-S(i\bm{\chi}^{\dagger}_{3}\times\bm{\chi}_{1},i\bm{\epsilon}_{2}\times\bm{\epsilon}_{4}^{*},\hat{r})T_\rho ]\notag\\
&-\frac{1}{8}\beta\beta_{S}g_{V}^{2}(\bm{\chi}^{\dagger}_{3}\cdot\bm{\chi}_{1})(\bm{\epsilon}_{4}^{*}\cdot\bm{\epsilon}_{2})Y_\omega\notag\\
&+\frac{1}{12}\lambda\lambda_{S}g_{V}^{2}[ 2(i\bm{\chi}^{\dagger}_{3}\times\bm{\chi}_{1})\cdot(i\bm{\epsilon}_{2}\times\bm{\epsilon}_{4}^{*})C_\omega\notag\\
&-S(i\bm{\chi}^{\dagger}_{3}\times\bm{\chi}_{1},i\bm{\epsilon}_{2}\times\bm{\epsilon}_{4}^{*},\hat{r})T_\omega ].
\end{align}

The effective potential of the $\Xi_{c}'\bar{D}$ and $\Xi_{c}\bar{D}^{*}$ coupled channels is:
\begin{align}
V_{\Xi_{c}'\bar{D}\rightarrow \Xi_{c}\bar{D}^{*}}&= -\frac{I_F gg_{4}}{6\sqrt{6}f_{\pi}^{2}}[\chi^{\dagger}_{3}\bm{\sigma}\cdot\bm{\epsilon}_{2}^{*}\chi_{1}C_\pi+\chi^{\dagger}_{3}S(\bm{\sigma},\bm{\epsilon}_{2}^{*},\hat{r})\chi_{1}T_\pi ]\notag\\
&-\frac{gg_{4}}{6\sqrt{6}f_{\pi}^{2}}[\chi^{\dagger}_{3}\bm{\sigma}\cdot\bm{\epsilon}_{2}^{*}\chi_{1}C_\eta+\chi^{\dagger}_{3}S(\bm{\sigma},\bm{\epsilon}_{2}^{*},\hat{r})\chi_{1}T_\eta ]\notag\\
&-\frac{I_F}{3\sqrt{6}}\lambda\lambda_{I}g_{V}^{2}[2\chi^{\dagger}_{3}\bm{\sigma}\cdot\bm{\epsilon}_{2}^{*}\chi_{1}C_\rho-\chi_{3}^{\dagger}S(\bm{\sigma},\bm{\epsilon}_{2}^{*},\hat{r})\chi_{1}T_\rho ]\notag\\
&-\frac{1}{3\sqrt{6}}\lambda\lambda_{I}g_{V}^{2}[2\chi^{\dagger}_{3}\bm{\sigma}\cdot\bm{\epsilon}_{2}^{*}\chi_{1}C_\omega-\chi_{3}^{\dagger}S(\bm{\sigma},\bm{\epsilon}_{2}^{*},\hat{r})\chi_{1}T_\omega ],
\end{align}
where the $I_F$ represents the isospin factor for specific isospin state and equals $-3$ or $1$ for $I=0$ or $I=1$ system, respectively.

\section{Inclusion of the $\Lambda\eta_c$ and $\Lambda J/\psi$ channels}\label{app:decay}
In this section, we construct the effective potentials for the $\Lambda\eta_c$ and $\Lambda J/\psi$ channels couples with the six channels in the main text, and then study the effects of these two channels on the predicted molecular states. The potentials are obtained from t-channels scattering amplitude in the same way as in Sec.~\ref{sec:2}. For the effective vertices $J/\psi D^{(*)}\bar D^{(*)}$ and $\eta_c D^{(*)}\bar D^{(*)}$, we employ the Lagrangian given in Ref.~\cite{Wang:2015xsa},
\begin{align}
    \mathcal{L}_{\mathcal{J}D^{(*)}\bar D^{(*)}}=&G_1 {\rm Tr}[\mathcal{J} \bar H^{\bar Q}i\overleftrightarrow{\partial}_\mu\gamma^\mu \bar H^{Q}]+h.c.,
\end{align}
where the $\mathcal{J}$ represents the spin doublet field in HQSS
\begin{align}
\mathcal{J} &= \frac{1+\slashed{v}}{2}\left( \left( J/\psi\right)^{\mu}\gamma_{\mu} - \eta_c\gamma_5 \right)\frac{1-\slashed{v}}{2}.
\end{align}
Following Ref~~\cite{Shimizu:2017xrg}, we adopt the value of the coupling constant $G_1=0.679~\rm{GeV}^{-3/2}$. On the other hand, accounting on the HQSS and SU(3) flavor symmetry, the effective Lagrangian including the $\Lambda D^{(*)}\Xi_c^{(',*)}$ interaction can be written as,     
\begin{align}
    \mathcal{L}_{B_8H^QS}=&G_2\sum_{\nu,\nu_1,\nu_2}\begin{pmatrix}8&8&1\\\nu&-\nu&0\end{pmatrix}\begin{pmatrix}3&6&8\\\nu_1&\nu_2&-\nu\end{pmatrix}\notag\\
    \times&(\bar B_8)_\nu \gamma_5\gamma^\mu (\bar H^{Q})_{\nu_1} (S_\mu)_{\nu_2}+h.c.,\label{eq:lagB8HS}\\
    \mathcal{L}_{B_8H^QB_{\bar 3}}=&G_3\sum_{\nu,\nu_1,\nu_2}\begin{pmatrix}8&8&1\\\nu&-\nu&0\end{pmatrix}\begin{pmatrix}3&3^*&8\\\nu_1&\nu_2&-\nu\end{pmatrix}\notag\\
    \times&(\bar B_{8})_{\nu}(\bar H^Q)_{\nu_1} (B_{\bar 3})_{\nu_2}+h.c.,\label{eq:lagB8HB3}
\end{align}
where $S_\mu$ and $\bar H^{Q}$ are heavy quark spin-doublet baryon and meson fields in HQSS as shown in Sec.~\ref{sec:2}, $B_8$ represents the SU(3) octet baryon fields. In the SU(3) flavor space, $S_\mu$, $B_8$ and $B_{\bar 3}$ belong to the $6$ and $8$ and $3^*$ baryon representations, while $\bar H^{Q}$ belongs to the $3$ meson representation. In this Lagrangian, the SU(3) flavor symmetry is preserved as the manner in Ref.~\cite{deSwart:1963pdg}, such that we extract the singlet from the product representation with help of SU(3) Clebsch-Gordan (CG) coefficient, which can be expressed interms of SU(2) CG coefficient multiplied by isoscalar factor,
\begin{align}
    \begin{pmatrix}R_1&R_2&R\\\nu_1&\nu_2&\nu\end{pmatrix}=\begin{pmatrix}R_1&R_2&R\\I_1Y_1&I_2Y_2&I Y
    \end{pmatrix}\mathcal{C}_{I_1I_{1z};I_2I_{2z}}^{II_z},
\end{align}
where parentheses at the right hand side denote the isoscalar factor, $\mathcal{C} _{I_1I_{1z};I_2I_{2z}}^{II_z}$ is the SU(2) CG coefficient, the isospin($I$) and its third component($I_z$) as well as the hyper-charge($Y$) quantum numbers of individual states in the eigenvalue diagram for representation $R$ are defined by $\nu=(I,I_z,Y)$. The relevant isoscalar factors are calculated with method in Ref.~\cite{deSwart:1963pdg} and listed in Tables~\ref{tab:isoscalfac36} and ~\ref{tab:isoscalfac33b}.  
\begin{table}[ht]
\caption{Isoscalar factor for the CG series $3\otimes 6=10\oplus 8$. Each entry with ``$\cdots$'' denotes that the isoscalar factor for such product representation is zero.}\label{tab:isoscalfac36}
\begin{ruledtabular}
\centering
\begin{tabular}{cc|cccc|cc}
$I$&$ Y$ & $I_1$&$ Y_1$ & $I_2$&$ Y_2$ & $10$ & $8$  \\
\hline
$3/2 $&$ 1$ & $1/2$&$ 1/3$ & $1 $&$ 2/3$ & $1$ &$\cdots$ \\
\hline
$1/2 $&$ 1$ & $1/2$&$ 1/3$ & $1 $&$ 2/3$ & $\cdots$ & $-1$ \\
\hline
$1 $&$ 0$ & $1/2$&$1/3$ & $1/2 $&$ -1/3$ & $\sqrt{2/3}$ & $\sqrt{1/3}$ \\
 & &$0$&$ -2/3$ & $1 $&$ 2/3$ & $\sqrt{1/3}$ & $-\sqrt{2/3}$ \\
\hline
$0 $&$ 0$ & $1/2$&$ 1/3$ & $1/2$&$ -1/3$ & $\cdots$ & $-1$ \\
\hline
$1/2 $&$ -1$ & $0$&$ -2/3$ & $1/2 $&$ -1/3$ & $\sqrt{2/3}$ & $-\sqrt{1/3}$ \\
 && $1/2$&$ 1/3$ & $0 $&$ -4/3$ & $\sqrt{1/3}$ & $\sqrt{2/3}$ \\
\hline
$0 $&$ -2$ & $0$&$ -2/3$ & $0 $&$ -4/3$ & $1$ & $\cdots$ \\
\end{tabular}
\end{ruledtabular}
\end{table} 
\begin{table}[ht]
\caption{Isoscalar factor for the CG series $3\otimes 3^*=8\oplus 1$. Each entry with ``$\cdots$'' denotes that the isoscalar factor for such product representation is zero.}\label{tab:isoscalfac33b}
\begin{ruledtabular}
\centering
\begin{tabular}{cc|cccc|cc}
$I$ & $Y$ & $I_1$ & $Y_1$ & $I_2$ & $Y_2$ & $8$ & $1$  \\
\hline
$1/2$ & $1$ & $1/2$ & $1/3$ & $0$ & $2/3$ & $1$ & $\cdots$ \\
\hline
$1$ & $0$ & $1/2$ & $1/3$ & $1/2$ & $-1/3$ & $1$ & $\cdots$ \\
\hline
$1/2$ & $-1$ & $0$ & $-2/3$ & $1/2$ & $-1/3$ & $1$ & $\cdots$ \\
\hline
$0$ & $0$ & $1/2$ & $1/3$ & $1/2$ & $-1/3$ & $\sqrt{1/3}$ & $\sqrt{2/3}$ \\
 & & $0$ & $-2/3$ & $0$ & $2/3$ & $\sqrt{2/3}$ & $-\sqrt{1/3}$ \\
\end{tabular}
\end{ruledtabular}
\end{table}
For the coupling constants, $g_{\Sigma_cDN}$ and $g_{\Lambda_cDN}$  are roughly estimated to be 2.69 and 13.5 in Refs.~\cite{Garzon:2015zva,Lu:2016nnt,Lin:2017mtz}. After expending the Lagrangian in Eq.~\eqref{eq:lagB8HS}and ~\eqref{eq:lagB8HB3} with these isoscalar factors, we can relate the coupling $G_2$ and $G_3$ with $g_{\Sigma_cDN}$ and $g_{\Lambda_cDN}$ as
\begin{align}
    G_2=\frac{1}{\sqrt{2m_D}}g_{\Sigma_cDN},~~~G_3=-\frac{1}{\sqrt{6m_D}}g_{\Lambda_cDN}.
\end{align}
With the Lagrangian above, we can derive the effective potentials for $\Lambda\eta_c\to \Xi_c^{(',*)}\bar D^{(*)}$ and $\Lambda J/\psi\to \Xi_c^{(',*)}\bar D^{(*)}$ transition the same method in Sec.\ref{sec:2}. They can be expressed in terms of the $Y_{\rm{ex}}$, $C_{\rm{ex}}$ and $T_{\rm{ex}}$ functions in the position space as
\begin{align}
    V_{1\to3}=&-\frac{\mathcal{B}^\prime m_{D^*}}{2}\chi^\dagger\chi q^0Y_{\rm{D^*}}+\frac{\mathcal{A} m_{D^*}}{4}\chi^\dagger\chi C_{\rm{D^*}},
\end{align}
\begin{align}
    V_{1\to4}&=\frac{\mathcal{B}m_{D^*}}{4\sqrt{3}}q^0\chi^\dagger\chi Y_{D^*}  +\frac{\mathcal{A}m_{D^*}}{24\sqrt{3}}\chi^\dagger\chi C_{D^*},
\end{align}
\begin{align}
    V_{1\to5}=&-\frac{\mathcal{A}^\prime m_{D}}{12}\chi^\dagger\left[\bm{\sigma}\cdot\bm{\epsilon}_2C_{D}+S(\bm{\sigma},\bm{\epsilon}_2,\hat{r})T_{D}\right]\chi\notag\\
    &-\frac{\mathcal{A}^\prime m_{D^*}}{12}\chi^\dagger\left[2\bm{\sigma}\cdot\bm{\epsilon}_2C_{D^*}-S(\bm{\sigma},\bm{\epsilon}_2,\hat{r})T_{D^*}\right]\chi,
\end{align}
\begin{align}
    V_{1\to7}&=\frac{\sqrt{3}\mathcal{A}m_{D}}{24}\left[\bm{\sigma}\cdot\bm{\epsilon}_2C_{D}+S(\bm{\sigma},\bm{\epsilon}_2,\hat{r})T_{D}  \right]\notag\\
    &+\frac{\mathcal{A}m_{D^*}}{12\sqrt{3}}\left[-2\bm{\sigma}\cdot\bm{\epsilon}_2C_{D^*}+S(\bm{\sigma},\bm{\epsilon}_2,\hat{r})T_{D^*}  \right],
\end{align}
\begin{align}
V_{1\to8}&=\frac{\mathcal{A}m_{D^*}}{6}\chi^\dagger\left[\bm{\sigma}\cdot(i\bm{\chi}\times\bm{\epsilon}_2)C_{D^*}+S(\bm{\sigma},(i\bm{\chi}\times\bm{\epsilon}_2),\hat{r})T_{D^*}  \right],
\end{align}
\begin{align}
    V_{2\to3}=&-\frac{\mathcal{A}^\prime m_{D}}{24}\chi^\dagger\left[\bm{\sigma}\cdot\bm{\epsilon}_4^*C_{D}+S(\bm{\sigma},\bm{\epsilon}^*_4,\hat{r})T_{D}  \right]\notag\\
    &+\frac{\mathcal{A}^\prime m_{D^*}}{6}\chi^\dagger\left[2\bm{\sigma}\cdot\bm{\epsilon}^*_4C_{D^*}-S(\bm{\sigma},\bm{\epsilon}^*_4,\hat{r})T_{D^*}  \right],
\end{align}
\begin{align}
    V_{2\to4}=&-\frac{\sqrt{3}\mathcal{A}m_{D}}{12}\chi^\dagger[\bm{\sigma}\cdot\bm{\epsilon}^*_4C_{D}+ S(\bm{\sigma},\bm{\epsilon}^*_4,\hat{r}) T_{D}]\notag\\
    &-\frac{\mathcal{A}m_{D^*}}{12\sqrt{3}}\chi^\dagger[-2\bm{\sigma}\cdot\bm{\epsilon}^*_4C_{D^*}+ S(\bm{\sigma},\bm{\epsilon}^*_4,\hat{r}) T_{D^*}  ],
\end{align}
\begin{align}
    V_{2\to5}=&\frac{\mathcal{A}^\prime m_{D}}{12}\chi^\dagger\left[\bm{\sigma}\cdot(i\bm{\epsilon}^*_4\times\bm{\epsilon}_2)C_{D}+S(\bm{\sigma},(i\bm{\epsilon}^*_4\times\bm{\epsilon}_2),\hat{r})T_{D} \right]\notag\\
    &+\frac{\mathcal{B}^\prime m_{D^*}}{4}\chi^\dagger\chi q^0\bm{\epsilon}^*_4\cdot\bm{\epsilon}_2 Y_{D^*}-\frac{\mathcal{A} m_{D^*}}{24}\chi^\dagger[ 3\bm{\epsilon}^*_4\cdot\bm{\epsilon}_2C_{D^*}\notag\\
    &-2\bm{\sigma}\cdot(i\bm{\epsilon}^*_4\times\bm{\epsilon}_2)C_{D^*}+S(\bm{\sigma},(i\bm{\epsilon}^*_4\times\bm{\epsilon}_2),\hat{r})T_{D^*}]
\end{align}
\begin{align}
    V_{2\to6}=-\frac{\mathcal{A}m_{D^*}}{6}\chi^\dagger\left[ \bm{\sigma}\cdot(i\bm{\chi}\times\bm{\epsilon}^*_4)C_{D^*}+ S(\bm{\sigma},(i\bm{\chi}\times\bm{\epsilon}^*_4),\hat{r})T_{D^*} \right],
\end{align}
\begin{align}
    V_{2\to 7 }=&-\frac{\mathcal{A}m_{D}}{4\sqrt{3}}\left[\bm\sigma\cdot(i \bm\epsilon^*_4\times\bm\epsilon_2)C_{D}+S(\bm\sigma,(i\bm\epsilon^*_4\times\bm\epsilon_2),\hat{r})T_{D}\right]\notag\\
&+\frac{\mathcal{B}m_{D^*}}{2\sqrt3}\chi^\dag\chi q^0 \bm\epsilon_4^*\cdot \bm\epsilon_2 Y_{D^*}-\frac{\mathcal{A}m_{D^*}}{12\sqrt{3}} \chi^\dag\Bigg\{3\bm\epsilon^*_4\cdot\bm\epsilon_2C_{D^*}\notag\\
&-2\bm\sigma\cdot(i \bm\epsilon^*_4\times\bm\epsilon_2)C_{D^*}+S(\bm\sigma,(i\bm\epsilon^*_4\times\bm\epsilon_2),\hat{r})T_{D^*}\Bigg\}\chi, 
\end{align}
\begin{align}
    V_{2\to 8 }=&-\frac{\mathcal{A}m_{D^*}}{6}\Bigg\{[\chi^\dagger\bm{\sigma}\cdot\bm{\chi} \bm{\epsilon}^*_4\cdot\bm{\epsilon}_2 +\chi^\dagger\bm{\sigma}\cdot\bm{\epsilon}_2\bm{\chi}\cdot\bm{\epsilon}^*_4\notag\\
&-\chi^\dagger\bm{\sigma}\cdot\bm{\epsilon}^*_4\bm{\chi}\cdot\bm{\epsilon}_2]C_{D^*}
    +[\chi^\dagger S(\bm{\sigma},\bm{\epsilon}_2,\hat{r}) \bm{\epsilon}^*_4\cdot\bm{\epsilon}_2 \notag\\&+\chi^\dagger S(\bm{\sigma},\bm{\epsilon}_2,\hat{r})\bm{\chi}\cdot\bm{\epsilon}^*_4
    -\chi^\dagger S(\bm{\sigma},\bm{\epsilon}^*_4,\hat{r})\bm{\chi}\cdot\bm{\epsilon}_2]T_{D^*}\Bigg\},
\end{align}
where $\mathcal{A}=-G_1G_2/\sqrt{m_\Lambda(E_\Lambda+m_\Lambda)}$, $\mathcal{A}^\prime=-G_1G_3/\sqrt{m_\Lambda(E_\Lambda+m_\Lambda)}$, $\mathcal{B}=-G_1G_2\sqrt{(E_\Lambda+m_\Lambda)/m_\Lambda}$, $\mathcal{B}^\prime=-G_1G_3\sqrt{(E_\Lambda+m_\Lambda)/m_\Lambda}$. We enumerate the ten channels of $\Lambda\eta_c$, $\Lambda J/\psi$, $\Xi_{c}\bar{D}$, $\Xi_{c}'\bar{D}$, $\Xi_{c}\bar{D}^{*}$, $\Xi_{c}^{*}\bar{D}$, $\Xi_{c}'\bar{D}^{*}$ and $\Xi_{c}^{*}\bar{D}^{*}$ with 1 to 8, and the $V_{i\to j}$ with subindex represents the potential corresponding to the transition from $i$th channel to $j$th channel.   
\bibliographystyle{apsrev4-1}
\bibliography{citebib.bib}
\end{document}